 \documentclass[final,5p,times,twocolumn,authoryear]{elsarticle}

\usepackage{amssymb}
\usepackage{txfonts}
\usepackage{pdflscape}
\usepackage{enumitem}

\usepackage{natbib}
\usepackage{graphicx}	
\usepackage{amsmath}	
\usepackage{amssymb}	
\usepackage{rotating}
\usepackage{listings}

\usepackage[T1]{fontenc}
\usepackage{ae,aecompl}

\journal{High Energy Astrophysics}

\begin{document}

\begin{frontmatter}



\title{The challenge of identifying \textit{INTEGRAL} sources on the Galactic plane}


\author[1]{Raffaella Landi}
\ead{raffaella.landi@inaf.it}
\affiliation[1]{organization={INAF -- Osservatorio di Astrofisica e Scienza dello Spazio},
            addressline={Via Piero Gobetti 93/3}, 
            city={Bologna},
            postcode={40129}, 
            state={Italy}}
    \author[1]{Loredana Bassani}
    \ead{loredana.bassani@inaf.it}

     \author[2]{Gabriele Bruni}
    \ead{gabriele.bruni@inaf.it}

\affiliation[2]{organization={INAF -- Istituto di Astrofisica e Planetologia Spaziali},
            addressline={Via del Fosso del Cavaliere 100}, 
            city={Roma},
            postcode={00133}, 
            state={Italy}}

\affiliation[3]{organization={INAF -- Istituto di Astrofisica Spaziale e Fisica Cosmica},
            addressline={Via Alfonso Corti 12}, 
            city={Milano},
            postcode={20133}, 
            state={Italy}}

    \author[3]{Manuela Molina}
    \ead{manuela.molina@inaf.it}

        \affiliation[4]{organization={Instituto de Astrof\'isica, Facultad de Ciencias Exactas, 
Universidad Andr\'es Bello}, 
           addressline={Fern\'andez Concha 700, Las Condes}, 
            city={Santiago RM},
            state={Chile}}

 \author[1,4]{Nicola Masetti}
    \ead{nicola.masetti@inaf.it}
    
 \author[1]{Angela Malizia}
    \ead{angela.malizia@inaf.it}
    
 \author[2]{Mariateresa Fiocchi}
    \ead{mariateresa.fiocchi@inaf.it}
    
 \author[2]{Angela Bazzano}
    \ead{angela.bazzano@inaf.it}

 \author[2]{Pietro Ubertini}
    \ead{pietro.ubertini@inaf.it}

\begin{abstract} 
The International Gamma-ray Astrophysics Laboratory 
(\emph{INTEGRAL}) has been surveying the sky above 20 keV since its launch in 2002 providing new 
insights into the nature of the sources that populate our Universe at soft $\gamma$-ray energies. 
The latest IBIS/ISGRI survey lists 929 hard X-ray sources, of which 113 are reported as 
unidentified, i.e. lacking a lower energy counterpart or simply not studied in other wavebands. 
To overcome this lack of information, we either browsed the X-ray archives, or, if no data in the 
X-ray band were available, we requested Target of Opportunity (ToO) observations with the X-ray 
Telescope (XRT) on-board the Neil Gehrels \emph{Swift} Observatory. Following this approach, we 
selected a sample of 10 objects for which X-ray data were key to investigate their nature. We 
found a single X-ray association for all of the sources, except for IGR J16267$-$3303, for which 
two X-ray detections were spotted within the IBIS positional uncertainty. We then browsed 
multi-waveband archives to search for counterparts to these X-ray detections at other wavelengths 
and analysed X-ray spectral properties to determine their nature and association with the 
high-energy emitter. As a result of our analysis, we identified the most likely counterpart for 7 
sources, although in some cases its nature/class could not be definitely assessed on the basis of 
the information collected. Interestingly, SWIFT J2221.6$+$5952, first reported in the
105-month \emph{Swift}/Burst Alert Telescope (BAT) survey, is the only source of the sample 
for which we did not find any counterpart at radio/optical/IR wavebands. Finally, we found that 
two IBIS source, IGR J17449$-$3037 and IGR J17596$-$2315 are positionally associated with a 
\emph{Fermi} Large Area Telescope (LAT) object. 
\end{abstract}




\end{frontmatter}




\section{Introduction}
\label{introduction}

In the last decade, high-energy instruments with imaging capabilities like \emph{INTEGRAL}/IBIS 
\citep{Ubertini2003} and \emph{Swift}/BAT \citep{Barthelmy2005} have revolutionised our view of 
the 20--100 keV sky by providing the deepest surveys of the entire celestial sphere. Overall, 
around 2000 sources have been discovered so far, a large part of which were new detections, 
either persistent or transient and their number is continuously increasing due to the constant 
accumulation of exposure. Many of these high-energy emitters have been optically identified 
thanks to X-ray follow-up observations that allow one to pinpoint, within the much smaller 
positional uncertainty, a single (rarely a double) optical/infrared (IR) counterpart, which is 
then classified through optical spectroscopy (see, e.g., \citealt{Koss2017}, 
\citealt{Karasev2018}, \citealt{Butler2009}, \citealt{Masetti2010}, \citealt{Masetti2012}, 
\citealt{Masetti2013}). Despite this large effort from the astronomical community over the years, 
a significant sample of sources are still unidentified/unclassified mainly due to their location 
on the Galactic plane, where identification and classification procedures are much more 
difficult. This is mainly due to the Galactic plane being more optically/IR crowded and also more 
reddened than the extragalactic sky. Crowdedness makes the identification of a single/double 
optical/IR counterpart difficult, as many objects may fall within the few arcseconds accuracy of 
the X-ray localisation, while reddening makes optical and often IR follow-up observations more 
challenging. Variability is another issue since objects seen occasionally by high-energy 
instruments may not be bright enough during follow-up at lower energies. For these reasons, a 
multi-wavelength approach, coupled to the most stringent positional accuracy available, is the 
key to deal with unidentified/unclassified IBIS and BAT sources lying along the Galactic plane. 
Sometimes, however, even this procedure is not sufficient to pinpoint a single counterpart and/or 
to understand the source nature/class, making the identification of these objects even more 
challenging.

Bearing this in mind, we focused our attention on those sources, listed in the latest IBIS/ISGRI 
survey \citep{Krivonos2022}, which are located on the Galactic plane and are still unidentified. 
Firstly, our approach was to browse X-ray archives in search of data at lower energies that can 
help in identifying the X-ray and hence the optical/IR counterpart; furthermore, information in 
the X-ray band allows the characterisation of these sources in terms of spectral shape, flux, 
absorption properties and variability. This approach allowed us to single out a set of 10 
high-energy sources located at $|b| \lesssim10^{\circ}$. For three of them (IGR J16267$-$3303, 
IGR J18006$-$3426, and IGR J19193$+$0754), we also applied for XRT ToO observations, which were 
executed. The aim of these requests was to obtain on-source observations which were not available 
in the archives.

We then made use of lower energy X-ray observations (\emph{Swift}/XRT, \emph{XMM-Newton}, and 
\emph{Chandra}) to provide the best available position, some variability information and X-ray 
spectral parameters. The best position of the optical/IR counterpart, thus pinpointed, was then 
searched on available catalogues to provide insights into the source nature and a tentative 
classification. As a result, we were able to spot the likely counterpart for 7 out of 10 objects; 
in the case of IGR J17327$-$4405, IGR J18006$-$3426, and IGR J19193$+$0754 the enhanced X-ray 
positional uncertainties are still too large to pinpoint a unique optical/IR association. 
Intriguingly, SWIFT J2221.6$+$5952 is the only source of the sample for which we did not find any 
counterpart at any other wavelength investigated; with this respect, being the source located in 
a crowded region of the Galactic plane, we encourage further multi-band follow-up observation to 
shed light on the nature of this high-energy emitter.

As a by-product of this analysis, we found that 4 \emph{INTEGRAL} objects are close to an 
unclassified \emph{Fermi}/LAT source and that two of them (IGR J17449$-$3037 and IGR 
J17596$-$2315) have probably GeV associations.

The paper is structured as follows: in Sect. 2 we briefly present the method adopted for the 
\emph{Swift}/XRT and \emph{XMM-Newton} data reduction and the criteria assumed for the spectral 
analysis. Sect. 3 is devoted to the discussion of the results for each source of our sample, 
making use of multi-frequency information. Conclusions are drawn in Sect. 4.

\section{Data reduction and analysis}
To characterise the behaviour of the selected sample in X-rays, we used data acquired with the 
X-ray Telescope (XRT, 0.3--10 keV, \citealt{Burrows2005}) on board the Neil Gehrels \emph{Swift} 
Observatory and with the \emph{XMM-Newton} satellite. The details of the observations are 
reported in the Appendix A (see Table~\ref{tabA1}).

To produce XRT screened event file we made use of the standard data pipeline package ({\sc 
xrtpipeline} v. 0.13.7). All data were extracted only in the Photon Counting (PC) mode 
\citep{Hill2004}. Source events were extracted within a circular region with a radius of 20 
pixels (1 pixel corresponding to 2.36 arcseconds) centred on the source position, while 
background events were extracted from a source-free region nearby the X-ray source of 
interest. To search for X-ray detections both within the 90\% and 99\% IBIS error circles, we 
analysed, by means of {\sc XIMAGE} v. 4.5.1, the 0.3--10 keV image of each observation and, then, 
estimated the X-ray position using the task {\sc xrtcentroid v. 0.2.9}.
In most cases we smoothed the X-ray images to visualise better the X-ray counterparts;
the presence of grains/features inside the XRT field of view (FoV) are indisputably spurious.

The source spectra were then extracted from the corresponding event file using the {\sc XSELECT} 
v. 2.5a software and generally binned using {\sc grppha} in an appropriate way, so that the 
$\chi^{2}$ statistic could be applied. For sources with fewer counts (typically around 50--60), 
data were binned to have at least 5 count per energy bin and the Cash statistic \citep{Cash1979} 
was adopted.

We used version v. 014 of the response matrices and created the individual ancillary response 
file using the task {\sc xrtmkarf} v. 0.6.4.

Data from the European Photo Imaging Cameras (EPIC-pn, 0.1--12 keV, \citealt{Turner2001}) were 
reprocessed using the \emph{XMM-Newton} Standard Analysis Software (SAS) version 20.0.0 and 
employing the latest available calibration files. Only patterns corresponding to single and 
double events (PATTERN$<$4) were taken into account and the standard selection filter FLAG = 0 
was applied. EPIC-pn nominal exposures were filtered for periods of high background, resulting in 
the cleaned exposures reported in Table~\ref{tabA1}. Source counts were extracted from a circular 
region of typically 20--30 arcseconds radius centred on the source (depending on the source 
brightness), while background spectra were extracted from two circular regions of 20 arcseconds 
radius each in source-free areas. The ancillary response matrices (ARF) and the detector response 
matrices (RMF) were generated using the XMM-SAS tasks {\sc arfgen} and {\sc rmfgen} and spectral 
channels were rebinned in order to achieve a minimum of 20 counts per bin. For EPIC observations 
we also adopted the same approach as for XRT to search for likely counterparts.

Table~\ref{tab1} lists all of the 10 IBIS sources analysed here together with 
their position and significance in $\sigma$ units as listed in \citet{Krivonos2022}; following 
\citet{Krivonos2007}, we have used a 4.2 and 3.0 arcminutes search radius at 90\% confidence 
level (c.l.) for a source detected at 5--6$\sigma$ and around 10$\sigma$ c.l., respectively.
For each of these $\gamma$--ray emitters, we then report the position and relative uncertainties 
(at 90\% c.l.) of the likely counterpart/s detected by XRT and \emph{XMM-Newton} within the 90\% 
and 99\% IBIS error circles, as well as the count rate in the 0.3--10 keV and 3--10 keV energy 
bands, and their angular distance from the \emph{INTEGRAL} position.
The X-ray counterparts highlighted in the XRT and \emph{XMM-Newton} images are labelled as reported in 
Table~\ref{tab1}. 

The spectral analysis was 
performed by means of the XSPEC package (v. 12.12.1, 
\citealt{Arnaud1996}); errors are quoted at 90\% c.l. for one parameter of interest 
($\Delta\chi^{2}$ = 2.71).

For objects with more than one pointing, we analysed each single observation and discussed the 
presence of source variability in each dedicated section. When the quality of the data was not 
good enough to perform a reliable spectral analysis, to improve the signal-to-noise ratio, we 
summed together all the available observations and analysed the average spectra. In the first 
instance, we adopted, as our basic model, a simple power law passing through Galactic absorption 
\citep{Kalberla2005}. If this baseline model was not sufficient to fit the data, we then 
introduced extra spectral components as required. 
The results of this analysis are shown in Table~\ref{tab2} and Table~\ref{tab3} where we report 
the best-fit parameters obtained by fitting XRT and \emph{XMM-Newton} data, respectively.
A more detailed description of the X-ray spectral analysis 
results is given in a dedicated section for each source.

\section{Notes on individual sources}

In the following, we present results on each individual source in the X-ray, optical, IR, and
radio (see Table~\ref{tabB1} in the Appendix B) band, 
discussing the overall properties found for each candidate counterpart.

\subsection{IGR J16173$-$5023}

IGR J16173$-$5023 was first analysed by \citet{Tomsick2012}, who used a \emph{Chandra} 
observation performed on October 11, 2011 to search for the X-ray counterpart of this high-energy 
emitter. They found two possible associations: a brighter one with a 2--10 keV flux of 
$5.87\times10^{-12}$ erg cm$^{-2}$ s$^{-1}$ and a dimmer one with a 2--10 keV flux of 
$0.5\times10^{-12}$ erg cm$^{-2}$ s$^{-1}$. No clear optical/IR counterpart could be found for 
the brighter source, but the detection of an iron line at 6.8 keV suggested the possibility that 
this might be a Cataclysmic Variable (CV) in the Norma region.

\begin{figure}
	\centering 
	\includegraphics[width=0.5\textwidth, angle=0]{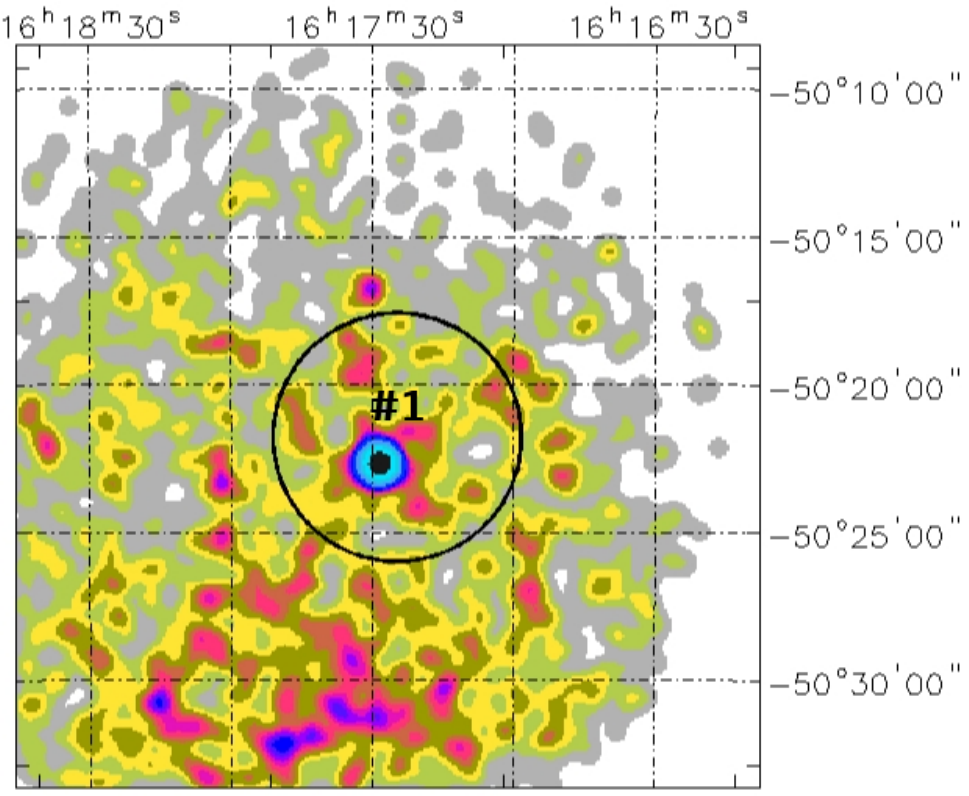}	
	\caption{XRT 0.3--10 keV image of the region surrounding IGR J16173$-$5023. The only XRT 
detection lies within the 90\% IBIS positional uncertainty (black circle).}
\label{fig1}%
\end{figure}

We analyse here XRT observations performed in 2012, 2017 and 2019 (see Table~\ref{tabA1}). The 
X-ray data indicate the presence of a single source inside the IBIS 90\% positional uncertainty 
(see Figure~\ref{fig1}), which is detected at $\sim$12$\sigma$ c.l. over the 0.3--10 keV energy 
range (see Table~\ref{tab1}). The source average spectrum is well fitted with the basic model 
(photon index of $\sim$1.5 and a 2--10 keV flux of $\sim$$3\times10^{-12}$ erg cm$^{-2}$ 
s$^{-1}$) plus thermal emission (Raymond-Smith model in XSPEC, \citealt{RaymondSmith1977}) to 
account for the excess below 2 keV (see Table~\ref{tab2}).

The source location and associated positional uncertainty indicate that this is the brightest 
source seen by \emph{Chandra} and further suggest that this is likely the counterpart to the 
persistent \emph{INTEGRAL} object. The XRT pointings also indicate that, although persistent, 
this is also a variable emitter since the X-ray flux increased by a factor of 4 during XRT 
monitoring (from 2012 to 2019).

Using the \emph{Chandra} position (from \citealt{Tomsick2012}) and a more restrictive positional 
uncertainty (R.A.(J2000) = $16^{\rm h}17^{\rm m}28^{\rm s}.26$, Dec.(J2000) = 
$-$$50^\circ22^{\prime}42^{\prime \prime}.50$, error radius of 1 arcsecond) with respect the XRT 
one, it is possible to pinpoint a single optical/IR counterpart designated as source 
5935091920522976256 in the \emph{Gaia} DR3 catalogue \citep{Brown2021}; the latest release of the 
catalogue (\emph{Gaia} DR3 Part 1 Main source, 2022\footnote{available at 
http://vizier.nao.ac.jp/viz-bin/Cat?I/355.} and originally in \citealt{Brown2021}) provides the 
following apparent magnitudes: $G= 18.09$, $BP =18.64$, $RP=17.36$, and an absolute parallax of 
$0.494\pm0.140$. Inferring the distance from \emph{Gaia} parallax is not a trivial issue nor is 
it easy to evaluate reddening corrections for \emph{Gaia} magnitudes; for these reasons, we rely 
on the parameters provided by the same \emph{Gaia} DR3 catalogue (version 2022) which quote a 
source distance of 1630 pc, an absolute $G$ magnitude of 6.15 and a reddening corrected $BP-RP$ 
colour of 0.79.
We further notice that the source is considered variable in \emph{Gaia} magnitude with a typical 
amplitude of 0.14 mag in $G$, $RP$, and 0.2 mag in $BP$ \citep{Mowlavi2021}.

Using this information and referring to the study of \citet{Eyer2019} on variable stars, it is 
possible to locate the source in the colour-absolute magnitude diagram close to the region 
populated by CV, making IGR J16173$-$5023 a CV candidate as first suggested by 
\citet{Tomsick2012}. Going more in detail, we also compared the source location in the 
\emph{Gaia} colour-absolute magnitude diagram of \citet{Abril2020} with that of various types of 
CV, finding that it might be an Intermediate Polar (IP) of around 7-hour period. This is fully 
compatible with the detection of an iron line at 6.8 keV seen by \emph{Chandra} 
\citep{Tomsick2012}, as well as with the source detection by IBIS, being IPs one of the most 
likely hard X-ray emitters seen by \emph{INTEGRAL} \citep{Lutovinov2020,Landi2009}.
Moreover, based on the 17--60 keV flux reported by \citet{Krivonos2022} 
and the luminosity function of hard X-ray emitting CVs \citep{Suleimanov2022}, the  
luminosity of IGR J16173$-$5023 at 1.6 kpc turns out to be $1.6\times10^{33}$ erg
s$^{-1}$, which further strenghtens the IP classification.

We also notice that the source is an H$\alpha$ emitter with magnitude $17.17\pm0.01$ as inferred 
from the VST Photometric H$\alpha$ Survey of the Southern Galactic Plane and Bulge (VPHAS$+$, 
\citealt{Drew2014}); the source is also present in the VISTA Variables in the Via Lactea Survey 
(VVV, \citealt{Minniti2017}) with IR magnitudes $J=16.052\pm0.011$, $H=15.708\pm0.022$, and 
$Ks=15.164\pm0.030$.

We conclude that the source is a Galactic object at a distance of 1.6 kpc and very likely a CV of 
the IP type.

\subsection{IGR J16267$-$3303}

\begin{figure}
   \centering
 
  \includegraphics[width=0.5\textwidth, angle=0]{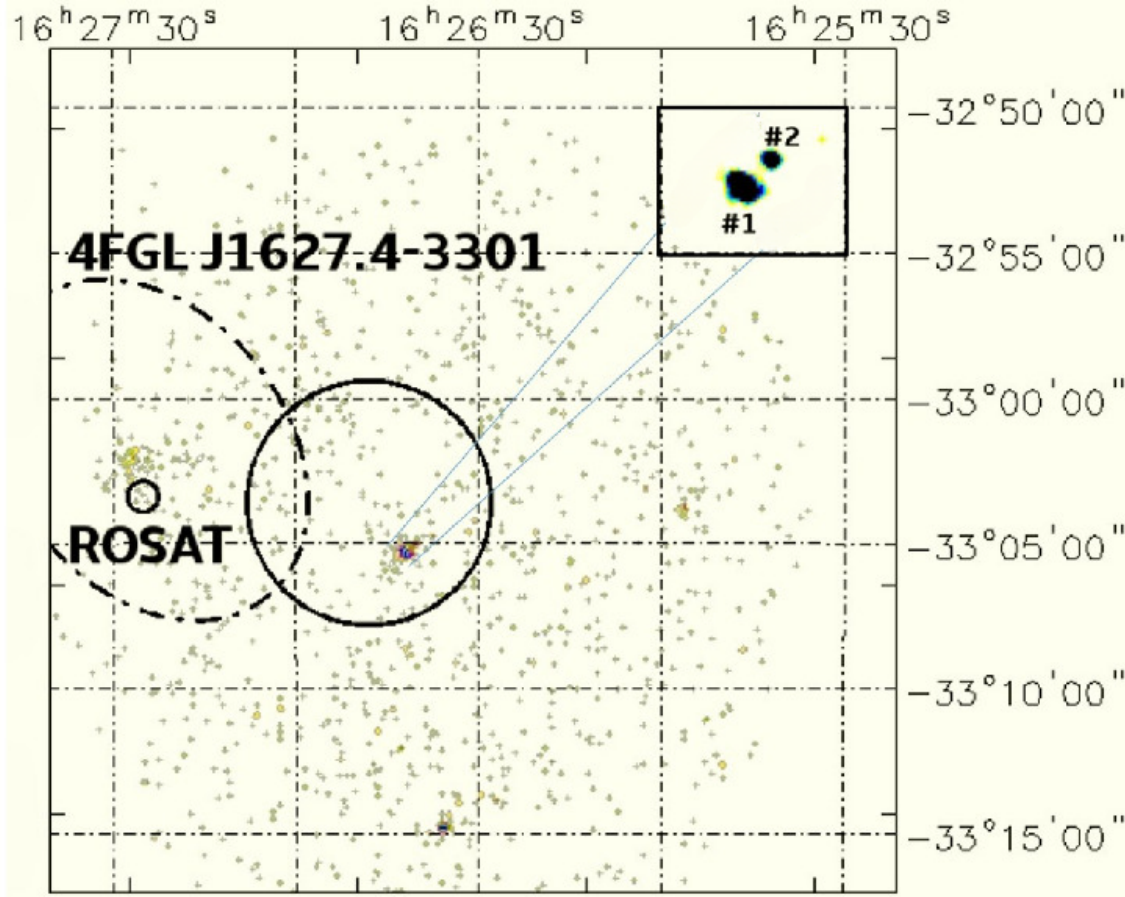}	
   \caption{XRT 0.3--10 keV image of the region surrounding IGR J16267$-$3303. Two X-ray sources 
(see close-up inset of the region) lie within the 90\% IBIS positional uncertainty (black 
circle). Also plotted are the error ellipse of the \emph{Fermi} source 4FGL J1627.4$-$3301 (black 
dash-dotted ellipse), which partially intersects the IBIS error circle, and the \emph{ROSAT} 
Faint source 1RXS J162725.1$-$330322 displayed with its positional uncertainty (smaller black 
circle).}
              \label{fig2}%
    \end{figure}

IGR J16267$-$3303 is a new \emph{INTEGRAL} source, listed for the first time in the 
\citet{Krivonos2022} catalogue. No X-ray observations had been available until we requested a
\emph{Swift}/XRT ToO observation performed on January 20, 2023 (see Table~\ref{tabA1}).
The source is interesting because of its proximity to a \emph{Fermi} object (4FGL 
J1627.4$-$3301), which was associated with a \emph{ROSAT} Faint source (1RXS 
J162725.1$-$330322\footnote{This source is located at: R.A.(J2000) = $16^{\rm h}27^{\rm m}25^{\rm 
s}.10$, Dec.(J2000) = $-$$33^\circ03^{\prime}22^{\prime \prime}.50$, error radius of 32 
arcseconds.}) and tentatively classified as a blazar of uncertain type \citep[and references 
therein]{Fan2022}. Figure~\ref{fig2} shows the XRT image of the sky region of 
interest with overlapping the IBIS and LAT positional uncertainties and the location of the 
\emph{ROSAT} source. Thanks to the positional accuracy of XRT, the position of the \emph{ROSAT} 
Faint object can be determined with a greater precision, namely at R.A.(J2000) = $16^{\rm 
h}27^{\rm m}26^{\rm s}.80$, Dec.(J2000) = $-$$33^\circ02^{\prime}00^{\prime \prime}.90$, with an 
associated uncertainty of 6 arcseconds. Unfortunately, this is still too large to identify a 
single optical/IR counterpart. We also note that, despite the low statistical quality of the data 
do not allow a full analysis, the \emph{ROSAT} source could be extended in X-rays.
However, this object lies outside the IBIS error circle, which leads us to exclude an association 
between the IBIS and \emph{Fermi} sources, although their positional uncertainties overlap 
slightly; we then focus, in the following, our attention only to the IBIS counterpart.

The XRT image indicates the presence of two X-ray detections inside the 90\% IBIS positional 
uncertainty (see zoomed section in Figure~\ref{fig2} and also Table~\ref{tab1}): source \#1 is 
the brightest and also hardest of the two and, therefore, the most likely counterpart. The X-ray 
average spectrum is of poor statistical quality to allow a full characterisation of the source 
behaviour, but our basic model provides either a flat spectrum ($\Gamma \sim -0.7$) and a 2--10 
keV flux of $\sim$$5\times10^{-12}$ erg cm$^{-2}$ s$^{-1}$ or an absorbed spectrum 
($N_{\rm{H(intr)}} \sim 4\times10^{22}$ cm$^{-2}$) if a photon index of 1.8 is assumed (see 
Table~\ref{tab2}).

    \begin{figure}
   \centering
   \includegraphics[width=0.5\textwidth, angle=0]{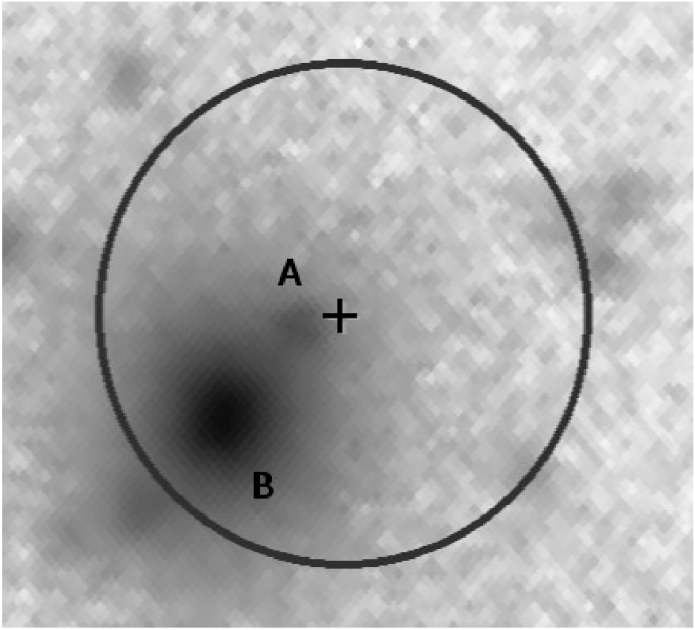}	

   \caption{DECaPS image of the region in the field of IGR J16267$-$3303 and centred around 
source \#1 located at R.A.(J2000)= $16^{\rm h}26^{\rm m}42^{\rm s}.05$ and Dec.(J2000) = 
$-$$33^\circ05^{\prime}19^{\prime \prime}.66$.
Within the XRT positional uncertainty (black circle, 6 arcsecond in radius) are clearly visible 
the two \emph{Gaia} sources labelled with A and B (see text).}

              \label{fig3}%

    \end{figure}

Figure~\ref{fig3} depicts the optical image, from the Dark Energy Camera Plane Survey 2 (DECaPS, 
\citealt{Saydjari2023}), of the region surrounding source \#1. Within the XRT error circle we 
find two objects (sources A and B in the Figure) which belong to the \emph{Gaia} DR3 catalogue 
(version 2022, originally in \citealt{Brown2021}). Source A, which is the closest source to the 
XRT best position (R.A.(J2000) = $16^{\rm h}26^{\rm m}42^{\rm s}.1351$, Dec.(J2000) = 
$-$$33^\circ05^{\prime}19^{\prime \prime}.937$), is dim ($G =21.00\pm0.03$) and difficult to 
classify. Source B (R.A.(J2000) = $16^{\rm h}26^{\rm m}42^{\rm s}.2762$, Dec.(J2000) = 
$-$$33^\circ05^{\prime}22^{\prime \prime}.117$) is further apart but, on the basis of its 
properties, easier to categorise as an active galaxy viewed through the Galactic plane: it is 
extended (see Figure~\ref{fig3}) and its \emph{WISE} (Wide-field Infrared Survey Explorer all sky 
survey, \citealt{Wright2010}) colours are typical of an AGN ($W1-W2=0.81$, $W2-W3=2.7$, 
$W3-W4=2.45$). Using these 3 colours and the 3-dimensional \emph{WISE} colour space diagram 
discussed by \citet{Dabrusco2019} (see their Figure 6) it is possible to characterise further the 
source as a blazar of the BL Lac type. The lack of radio emission in the source at 1 mJy level 
(see Table~\ref{tabB1}) is surprising but not unexpected \citep{Massaro2017}.

Source \#2 is dimmer (by a factor of 14) in X-rays and softer than object \#1, but it is still a 
possible extra association. Its average spectrum provides a 2--10 keV flux of 
$\sim$$2\times10^{-13}$ erg cm$^{-2}$ s$^{-1}$ by assuming the basic model with the photon index 
frozen to 1.8. Within the XRT positional uncertainty we find only one source which is also listed 
in the \emph{WISE} catalogue with colours once more typical of an AGN, and possibly of a BL Lac 
type blazar ($W1-W2=0.62$, $W2-W3=2.38$, $W2-W3$$\ge$3.27, \citealt{Dabrusco2019}).

Also this possible counterpart is not detected in radio at a flux limit of 1 mJy (see 
Table~\ref{tabB1}), thus making the \emph{WISE} colours the only signature of its AGN nature.

Based on the available information we suggest that source \#1 is the most likely counterparts to 
IGR J16267$-$3303, although a contribution from source \#2 is possible; in any case, this 
\emph{INTEGRAL} emitter is most likely an extragalactic source viewed through the Galactic plane.

\begin{table*}
\begin{center}
\caption{\emph{INTEGRAL}/IBIS position and source detection significance of the 10 selected 
sources. For each high-energy emitter, the objects detected by XRT or \emph{XMM-Newton}, within 
the 90\% IBIS positional uncertainties, are reported along with their relative count rates in the 
0.3--10 and 3--10 keV energy range, and their angular distance from the \emph{INTEGRAL} position. 
The XRT and \emph{XMM-Newton} error radii are given at $90\%$ confidence level.}
\label{tab1}
\footnotesize
\begin{tabular}{l c c c c c c c}
\hline
\hline
XRT source  &     R.A.     &     Dec.   &   error   &  \multicolumn{2}{c}{Count rate} &  Distance$^{a}$  &  Instr.  \\
            &              &            &          &   (0.3--10 keV)  &  (3--10 keV)  &    &   \\
  &   (J2000) &  (J2000) &   (arcsecond )  &  (10$^{-3}$ counts s$^{-1}$) & (10$^{-3}$ counts s$^{-1}$) &  (arcmin)  &  \\
\hline
\hline
  & &     &   &   &   &   &   \\
\multicolumn{7}{c}{\textbf{IGR J16173$-$5023 (R.A.(J2000) = $16^{\rm h}17^{\rm m}24^{\rm s}.72$,
Dec.(J2000) = $-$$50^\circ21^{\prime}46^{\prime \prime}.80$}, 7.4$\sigma$)}  \\
   & &     &   &   &   & \\
\#1  & $16^{\rm h}17^{\rm m}28^{\rm s}.37$ & $-$$50^\circ22^{\prime}40^{\prime \prime}.22$ & 3.84 & $41.18\pm3.30$ (12.5$\sigma$) &  $20.54\pm2.30$ (8.9$\sigma$)& 0.70 &  XRT \\
   & &     &   &   &  &  &   \\
   \hline
      & &     &   &   &   & &  \\
      \multicolumn{7}{c}{\textbf{IGR J16267$-$3303 (R.A.(J2000) = $16^{\rm h}26^{\rm m}48^{\rm s}.00$,
Dec.(J2000) = $-$$33^\circ03^{\prime}36^{\prime \prime}.00$}, 4.4$\sigma$)}  \\
  & &     &   &   &  & &   \\
\#1  & $16^{\rm h}26^{\rm m}42^{\rm s}.05$ & $-$$33^\circ05^{\prime}19^{\prime \prime}.66$ & 6.00 & $26.77\pm3.50$ (7.6$\sigma$) &  $20.37\pm3.00$ (6.8$\sigma$)& 2.42 & XRT\\
\#2  & $16^{\rm h}26^{\rm m}40^{\rm s}.41$ & $-$$33^\circ05^{\prime}00^{\prime \prime}.22$. & 6.00 &  $9.11\pm2.00$ (4.6$\sigma$) &  $3.91\pm1.30$ (3.0$\sigma$)& 2.35 & XRT\\
   & &     &   &   &  &  &  \\
   \hline
     & &     &   &   &   &  &  \\
   \multicolumn{7}{c}{\textbf{IGR J17315$-$3221 (R.A.(J2000) = $17^{\rm h}31^{\rm m}33^{\rm s}.60$,
Dec.(J2000) = $-$$32^\circ21^{\prime}36^{\prime \prime}.00$}, 4.2$\sigma$)}  \\
   & &     &   &   &  & &   \\
\#1  & $17^{\rm h}31^{\rm m}13^{\rm s}.85$ & $-$$32^\circ22^{\prime}05^{\prime \prime}.26$ & 3.88 & $21.20\pm2.10$ (10.1$\sigma$)&  $13.32\pm1.70$ (7.8$\sigma$)& 4.19 & XRT\\
   & &     &   &   &   &  &  \\
   \hline
      & &    &   &   &   &  &   \\
      \multicolumn{7}{c}{\textbf{IGR J17327$-$4405 (R.A.(J2000) = $17^{\rm h}32^{\rm m}45^{\rm s}.60$,
Dec.(J2000) = $-$$44^\circ06^{\prime}00^{\prime \prime}.00$}, 7.3$\sigma$)}  \\
  & &     &   &   &  &  \\
\#1  & $17^{\rm h}32^{\rm m}53^{\rm s}.05$ & $-$$44^\circ07^{\prime}29^{\prime \prime}.40$ & 3.66 & $47.52\pm2.70$ (17.6$\sigma$)&  $32.07\pm2.20$ (14.6$\sigma$)& 1.72 & XRT\\
   & &     &   &   &  &  &  \\
   \hline
      & &     &   &   &   &   &  \\
      \multicolumn{7}{c}{\textbf{IGR J17449$-$3037 (R.A.(J2000) = $17^{\rm h}44^{\rm m}55^{\rm s}.20$,
Dec.(J2000) = $-$$30^\circ37^{\prime}12^{\prime \prime}.00$}, 5.3$\sigma$)}  \\
   & &     &   &   &   &   &  \\
\#1 & $17^{\rm h}45^{\rm m}07^{\rm s}.97$ & $-$$30^\circ39^{\prime}05^{\prime \prime}.90$ & 1.1 & $53.38\pm3.22$ (16.6$\sigma$)&  $47.95\pm2.99$ (16.0$\sigma$)& 3.58 & XMM\\
   & &     &   &   &  &  &   \\
   \hline  
      & &     &   &   &   &  & \\
    \multicolumn{7}{c}{\textbf{IGR J17596$-$2315 (R.A.(J2000) = $17^{\rm h}59^{\rm m}38^{\rm s}.40$,
Dec.(J2000) = $-$$23^\circ16^{\prime}12^{\prime \prime}.00$}, 9.4$\sigma$)}  \\
  & &     &   &   &  & &  \\
\#1  & $17^{\rm h}59^{\rm m}46^{\rm s}.56$ & $-$$23^\circ13^{\prime}56^{\prime \prime}.43$ & 4.32 & $8.24\pm2.60$ (3.2$\sigma$)&  $5.28\pm1.90$ (2.8$\sigma$) & 3.62 & XRT \\
   & &     &   &   &  & &  \\
   \hline
      & &     &   &   &   &  & \\
      \multicolumn{7}{c}{\textbf{IGR J18006$-$3426 (R.A.(J2000) = $18^{\rm h}00^{\rm m}40^{\rm s}.80$,
Dec.(J2000) = $-$$34^\circ27^{\prime}00^{\prime \prime}.00$}, 5.0$\sigma$)}  \\
   & &     &   &   &  &  & \\
\#1  & $18^{\rm h}00^{\rm m}50^{\rm s}.55$ & $-$$34^\circ23^{\prime}20^{\prime \prime}.58$ & 4.10 & $27.31\pm3.80$ (7.2$\sigma$)&  $5.94\pm1.80$ (3.3$\sigma$)& 3.66 & XRT \\
   & &     &   &   &   &  &  \\
   \hline
      & &     &   &   &   & &  \\
   \multicolumn{7}{c}{\textbf{IGR J19071$+$0716 (R.A.(J2000) = $19^{\rm h}07^{\rm m}07^{\rm s}.20$,
Dec.(J2000) = $+$$07^\circ16^{\prime}12^{\prime \prime}.00$}, 4.2$\sigma$)}  \\
   & &     &   &   &   &  &  \\
\#1  & $19^{\rm h}07^{\rm m}06^{\rm s}.33$ & $+$$07^\circ20^{\prime}04^{\prime \prime}.58$ & 5.02 & $5.03\pm1.00$ (5.0$\sigma$)&  $2.84\pm0.77$ (3.7$\sigma$)& 4.15 & XRT\\
  & &     &   &   &   &   & \\
   \hline
      & &     &   &   &   &  &  \\
  \multicolumn{7}{c}{\textbf{IGR J19193$+$0754 (R.A.(J2000) = $19^{\rm h}19^{\rm m}16^{\rm s}.80$,
Dec.(J2000) = $+$$07^\circ54^{\prime}28^{\prime \prime}.80$}, 4.7$\sigma$)}  \\
   & &     &   &   &   &  &   \\
\#1  & $19^{\rm h}19^{\rm m}06^{\rm s}.57$ & $+$$07^\circ59^{\prime}28^{\prime \prime}.48$ & 6.21 & $13.52\pm3.50$ (3.9$\sigma$)&    -- & 5.25 & XRT\\
   & &     &   &   &   &  &  \\ 
   \hline
      & &     &   &   &   &  &   \\
  \multicolumn{7}{c}{\textbf{SWIFT J2221.6$+$5952 (R.A.(J2000) = $22^{\rm h}22^{\rm m}09^{\rm s}.60$,
Dec.(J2000) = $+$$59^\circ52^{\prime}48^{\prime \prime}.00$}, 5.2$\sigma$)}  \\
    & &     &   &   &   &  &   \\
\#1  & $22^{\rm h}21^{\rm m}59^{\rm s}.82$ & $+$$59^\circ51^{\prime}52^{\prime \prime}.33$ & 4.10 & $18.07\pm2.10$ (8.6$\sigma$)&   $8.82\pm1.50$ (5.9$\sigma$)& 1.81 & XRT\\
   & &     &   &   &   &  &  \\ 
   \hline
   \hline
\end{tabular}
\begin{list}{}{}
\item $^{a}$: Angular distance from the \emph{INTEGRAL} position. 
\end{list}
\end{center}
\end{table*}

\subsection{IGR J17315$-$3221}

This \emph{INTEGRAL} source was recently discussed by \citet{Ferrigno2022} (see their 
Appendix). Although previously classified in the literature as a possible supergiant X-ray 
binary, these authors suggested that it may be the product of a data analysis artifact and thus 
should be disregarded for future studies. They motivated this conclusion with the lack, within 2 
arcminutes from the IBIS position reported by \citet{Krivonos2012}, of an X-ray counterpart in 
the XRT image obtained by summing observations made over the period 2012--2016. They also 
reanalysed all \emph{INTEGRAL} archival data available around the position of IGR J17315$-$3221 
and found no detection at the source position. We note, however, that: a) the 2.2 arcminutes 
error radius provided by \citep{Krivonos2012} is only at the 68\% c.l. and b) that the source is 
further confirmed in the most recent catalogue of \citet{Krivonos2022}.

\begin{figure}
   \centering \includegraphics[width=0.5\textwidth, angle=0]{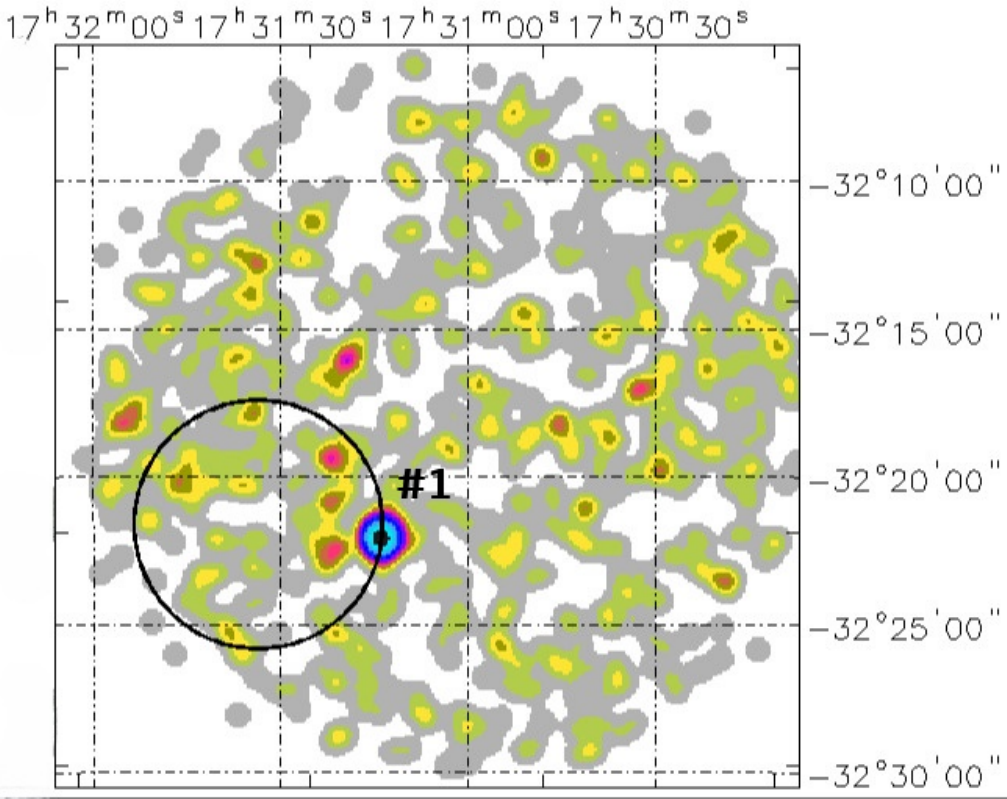}	
   \caption{XRT 0.3--10 keV image of the region surrounding IGR J17315$-$3221. XRT detects only 
one source located at the border of the 90\% IBIS positional uncertainty (black circle).}
  \label{fig4}%
    \end{figure}

We therefore reanalysed \emph{Swift}/XRT and \emph{XMM-Newton} observations of the source to 
search for an X-ray counterpart to this high-energy emitter within the 90\% of the position 
uncertainty associated with the most recent \emph{INTEGRAL} location. First, we analysed 
individual XRT pointings (performed on September 1, 2012 and during the period August 8--20, 
2016, see Table~\ref{tabA1}) to search for source variability. The lack of evidence for a 
significant flux variation led us to add all observations together to improve the signal-to-noise 
ratio and pinpoint the most likely counterpart. In Figure~\ref{fig4}, we show the XRT image,
in the 0.3--10 keV energy band,
of the region surrounding IGR J17315$-$3221, where it is evident the presence of a source 
at the border of the 90\% \emph{IBIS} error circle. The source position and relative error are 
listed in Table~\ref{tab1}. The average XRT spectrum, modelled with our basic model, provides a 
flat photon index ($\Gamma \sim 0.7$) and a 2--10 keV flux of $\sim$$2\times10^{-12}$ erg 
cm$^{-2}$ s$^{-1}$ (see Table~\ref{tab2}).

Then, we browsed the \emph{Chandra} Source Catalogue 2.1\footnote{available at: 
https://asc.harvard.edu/csc/about2.1.html.}, finding that the XRT source was present (2CXO 
J173113.7$-$322204, detected at $\sim$$7\sigma$ c.l.), albeit slightly dimmer (by a factor around 
1.8). The best \emph{Chandra} position is at R.A.(J2000) = $17^{\rm h}31^{\rm m}13^{\rm s}.68$, 
Dec.(J2000) = $-$$32^\circ22^{\prime}04^{\prime \prime}.80$ (error radius of 0.76 arcseconds), 
while the 0.5--7 keV flux is $(7.10\pm1.00)\times10^{-13}$ erg cm$^{-2}$ s$^{-1}$.

We also turned our attention to the \emph{XMM-Newton} data, as we find that a pointing was 
recently performed towards this region on September 14, 2022 (see Table~\ref{tabA1}). The 
analysis of the \emph{XMM-Newton} data provides a further detection of the source at a position 
compatible with that of XRT and \emph{Chandra}. This observation, moreover, yields the best yet 
available X-ray spectrum of the source. Our basic model applied to the \emph{XMM-Newton} data 
shows residuals hinting at an excess around 6 keV. The addition of an iron line component, with 
the line width $\sigma$ fixed to 0.01 keV, is required only at 93\% ($\Delta\chi^{2}/\nu = 6/2$), 
yielding a line centroid $E_{\rm C} \sim 6.4$ keV and an Equivalent Width ($EW$) around 208 eV; the 
0.2--12 keV flux is $\sim$$1.9\times10^{-12}$ erg cm$^{-2}$ s$^{-1}$ (see Table~\ref{tab3}).

Moreover, we note that the source is reported also as an \emph{XMM-Newton} Slew Survey 
\citep{Saxton2008} detection with a 0.2--12 keV flux of $(1.7\pm0.6)\times10^{-12}$ erg cm$^{-2}$ 
s$^{-1}$ in a slew made on August 29, 2014. Since the source was monitored by various X-ray 
detectors over a quite large period of years, we can infer a flux variability by a factor up to 
1.8 from 2016 to 2022.

Thanks to the restricted source location obtained by \emph{Chandra}, we are also able to pinpoint 
a possible optical/IR counterpart to the source, even if it is located in a crowded region of the 
Galactic plane (Scorpius region). Within the \emph{Chandra} positional uncertainty there is one 
single object reported in the \emph{Gaia} DR3 catalogue as source 4054812795194499968 
\citep{Brown2021}; it has an apparent $G$ magnitude of $20.68\pm0.01$ and a $BP-RP$ colour index 
of 1.90. Its distance can be found in the catalogue of \citet{Bailer-Jones2018} with a value of 
$4.33^{+2.14}_{-2.19}$ kpc. Adopting this distance, we obtain an absolute $G$ magnitude of 7.5 
(unfortunately we could not find a reliable reddening correction for this source); using this and 
the $BP-RP$ colour of 1.90, we locate the source in the \emph{Gaia} colour-absolute magnitude 
diagram close to the binary sequence zone \citep{Eyer2019}. The \emph{Gaia} object is also 
present in the VPHAS$+$ catalogue \citep{Drew2014} with an H$\alpha$ magnitude of $20.59\pm0.13$.

However, at near-IR frequencies the situation is more complex as the VVV catalogue 
\citep{Minniti2017} reports two possible counterparts: the \emph{Gaia} object (VVV 
J173113.69$-$322204.99) and another source (VVV J173113.64$-$322205.28) nearby, so for the time 
being we cannot pinpoint the actual optical/IR counterpart for this source.

We suggest in any case that this is likely a variable Galactic source, possibly of binary nature.

\subsection{IGR J17327$-$4405/SWIFT J1732.6$-$4408}

\begin{figure}
   \centering
    \includegraphics[width=0.5\textwidth, angle=0]{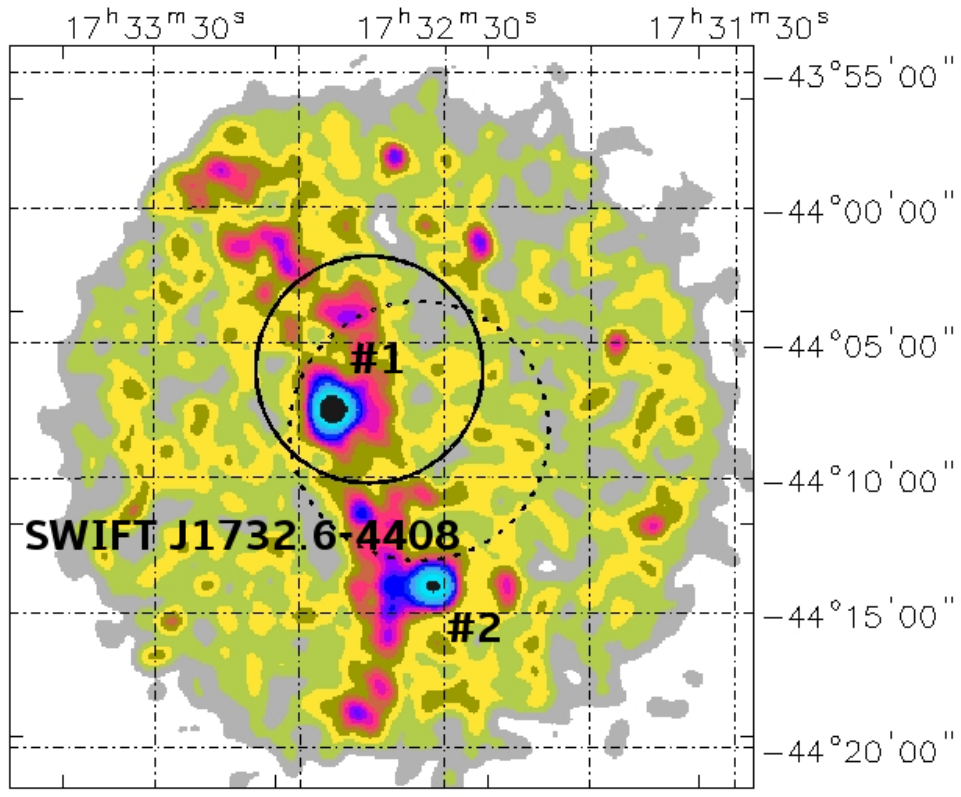}	

   \caption{XRT 0.3--10 keV image of the region surrounding IGR J17327$-$4405. XRT detects a 
source within the 90\% IBIS error circle (black circle), which also lies inside the positional 
uncertainty (black-dotted circle) of the \emph{Swift}/BAT source SWIFT J1732.6$-$4408. Source \#2 
is a transient object detected in the February 24, 2019 pointing (see text).}
              \label{fig5}%
    \end{figure}

This source is also reported in recent \emph{Swift}/BAT catalogues (see for example the 105- and 
the 157-month surveys\footnote{available at: https://swift.gsfc.nasa.gov/results/bs157mon/}, 
\citealt{Oh2018}), where it is listed as a source of unknown class and with no X-ray counterpart. 
However, this sky region was observed by XRT on a number of occasions over the 2017--2021 
period, although most of these pointings are of low exposure (see Table~\ref{tabA1}). The 0.3--10 
keV image of the region surrounding IGR J17327$-$4405, obtained stacking together all the XRT 
pointings, is displayed in Figure~\ref{fig5}: two sources are clearly detected, although only one 
falls within the positional uncertainty of both IBIS and BAT. Since the statistics on individual 
pointing is quite poor and the flux variation not so large (by a factor around 1.2), we decided 
to sum all the observations together and perform the spectral analysis of the average data.

The spectrum of source \#1, well described by our basic model, shows a flat spectral index 
($\Gamma \sim 0$), and requires an additional thermal component (MEKAL in {\sc XSPEC}; 
\citealt{Mewe1985}) with a $kT \sim 0.14$ keV; the 2--10 keV flux is $\sim$$6.7\times10^{-12}$ 
erg cm$^{-2}$ s$^{-1}$ (see Table~\ref{tab2}).

Unfortunately, the XRT positional uncertainty, even if small, is insufficient to pinpoint a 
unique optical candidate. For example, in the \emph{Gaia} DR3 catalogue (version 2022) 
\citep{Brown2021} there are three objects within the X-ray positional uncertainty, all of which 
are likely Galactic with distances in the range 2--6 kpc; one of these is also variable in 
\emph{Gaia}, suggesting a possible association with the XRT source. This ambiguity can only be 
resolved by means of a \emph{Chandra} pointing, which should be able to pinpoint a unique 
association.

The other source (\#2 in Figure~\ref{fig5}), is also an interesting object as it is extremely 
variable: it was quite bright on February 27, 2019 reaching a flux of $\sim$$1.7\times10^{-11}$ 
erg cm$^{-2}$ s$^{-1}$. However, the source was barely detected on February 1, 2019 with a 2--10 
keV flux $\sim$$1.4\times10^{-12}$ erg cm$^{-2}$ s$^{-1}$; no detection was found in the previous 
and following pointings.

The source is located at R.A.(J2000) = $17^{\rm h}32^{\rm m}32^{\rm s}.55$, Dec.(J2000) = 
$-$$44^\circ14^{\prime}02^{\prime \prime}.10$, with an associated error of 3.8 arcseconds. It is 
detected at $\sim$16$\sigma$ and $\sim$7$\sigma$ c.l. in the 0.3--10 keV energy range and above 3 
keV, respectively. Our basic model (power law with $\Gamma = 1.30^{+0.58}_{-0.77}$) does not 
provide a good fit to the data as the residuals support the presence of an excess below 2 keV 
that can be modelled with a thermal bremsstrahlung with $kT = 0.30^{+0.20}_{-0.13}$ keV. 

Given the extremely variable behaviour of this object versus the persistent nature of the hard 
X-ray source \#1, as well as its location outside the positional uncertainty of both IBIS and 
BAT, we exclude any association between this and source \#2, but note that this XRT source is 
nevertheless an interesting discovery for its extremely variable behaviour.

We therefore conclude that source \#1 is the likely association and further
suggests it is a Galactic source.

\begin{table*}
\begin{center}
\caption{Data analysis results of the averaged \emph{Swift}/XRT spectra. Frozen parameters are 
written in square brackets; errors are given at the 90\% confidence level.}
\label{tab2}
\small
\begin{tabular}{lccccccc}
\hline
\hline
 Source & $N_{\rm H(Gal)}$ & $N_{\rm H(int)}$ & $kT$ &  $\Gamma$  &  $\chi^{2}/d.o.f.$ 
 &  $C-stat/d.o.f.$   &   $F_{\rm (2-10~keV)}$ \\
  & ($10^{22}$ cm$^{-2}$) &  ($10^{22}$ cm$^{-2}$)   & (keV)  &   & & & ($10^{-12}$ erg cm$^{-2}$ s$^{-1}$)\\
\hline
\hline
  &   &   & & & & &\\
\multicolumn{8}{c}{\textbf{IGR J16173$-$5023}}\\
   &   &   & & & & &  \\
\#1$^{a}$  &  2.13  &       --     &  $0.19^{+0.32}_{-0.11}$     &    $1.51^{+0.53}_{-0.59}$   & 6.1/11    &  --   &   $3.48\pm0.24$  \\
   &   &   &   & & &  &   \\
\hline
  &   &   & & & & & \\
  \multicolumn{8}{c}{\textbf{IGR J16267$-$3303}}\\
   &   &   & & &  &  & \\
\#1    &  0.158  &       --     & --      &    $-0.61^{+0.41}_{-0.51}$   & --    &  9.7/10  &
$4.71\pm0.61$ \\
       &  0.158  &  $3.79^{+2.59}_{-1.23}$ &     --   & [1.8]  &  --  &  &
       $2.14\pm0.27$ \\
\#2    &  0.158  &       --     & --      &  [1.8]  &   --  &  4.2/3   &    
$0.18\pm0.05$ \\
   &   &   &  & & & & \\
\hline
  &   &   & & &  & &\\
\multicolumn{8}{c}{\textbf{IGR J17315$-$3221}}\\
   &   &   & & & & &  \\
\#1        &  1.24   &    --       &        --       &    $0.67^{+0.36}_{-0.40}$  &  
15.7/10  &    --   & $2.18\pm0.18$ \\
  &   &   & & & &  &\\
\hline
 &   &   & & & & &\\
\multicolumn{8}{c}{\textbf{IGR J17327$-$4405}}\\
   &   &   & & & &   & \\
\#1$^{b}$   & 0.307  &     --       &  $0.14\pm0.08$ &    $-0.04\pm0.16$         & 
11.2/20   &  --   &  $6.72\pm0.32$ \\
 &   &   & & &   & &\\
\hline
  &   &   & & &   &  &\\
\multicolumn{8}{c}{\textbf{IGR J17596$-$2315}}\\
   &   &   &  &  & & & \\
\#1           & 1.23   &      --             &    --         & [1.8]  &  --  &  5.5/7  &
$0.86\pm0.17$ \\
  &   &   & & &  & & \\
\hline
 &   &   & & &  &  &\\
  \multicolumn{8}{c}{\textbf{IGR J18006$-$3426}}\\
   &   &   & & & & & \\
\#1        & 0.228   &            --            &        --         & $1.13^{+0.43}_{-0.46}$  & 
6.5/7     &  --   &   $3.11\pm0.34$  \\
 &   &   & & & &  & \\
\hline
&  & &  &   & &   & \\
\multicolumn{8}{c}{\textbf{IGR J19071$+$0716}}\\
   &   &   & & &  & &  \\
\#1        & 1.56   &            --            &        --         & $1.46^{+0.88}_{-0.96}$  & 
    --   &  5.7/5    &       $0.35\pm0.07$  \\
  &   &   & & & &  &\\
  \hline
 &   &   & & & & & \\
  \multicolumn{8}{c}{\textbf{IGR J19193$+$0754}}\\
   &   &   & & & & & \\
\#1$^{b}$        & 0.554   &            --            &    $0.60^{+0.52}_{-0.31}$      &  --    & 
--    & 2.9/2    &   $0.018\pm0.04$  \\
  &   &   & & &  & &\\
\hline
  &   &   & & & & & \\
  \multicolumn{8}{c}{\textbf{SWIFT J2221.6$+$5952}}\\
   &   &   & & & &  \\
\#1          & 0.828   &            --            &  --     &  $0.90^{+0.44}_{-0.47}$    & 
5.6/8    & --  &   $1.33\pm0.15$  \\
             & 0.828   & $0.81^{+0.81}_{-0.53}$   &  --     &  [1.8] &  8.2/8  & --  &   $0.89\pm0.10$  \\
 &   &   & & & &  &\\
\hline
\hline
\end{tabular}
\begin{list}{}{}
\item $^{a}$: In this case, the best-fit model requires a thermal component 
(Raymond-Smith model in {\sc XSPEC});
\item $^{b}$: For this source, the best fit model requires a thermal component (MEKAL in {\sc XSPEC}).
\end{list}
\end{center}
\end{table*}

\begin{table*}[t]
\begin{center}
\caption{\emph{XMM-Newton} spectral analysis results; errors are given at the 90\% confidence level.}
\label{tab3}
\small
\begin{tabular}{lccccc}
\hline
\hline
 Source & $N_{\rm H(Gal)}$ & $N_{\rm H(int)}$ &   $\Gamma$  &  $\chi^{2}/d.o.f.$ 
 &  $F_{\rm (0.2-12~keV)}$ \\
  & ($10^{22}$ cm$^{-2}$) &  ($10^{22}$ cm$^{-2}$)   &    &  & ($10^{-12}$ erg cm$^{-2}$ s$^{-1}$)\\
\hline
\hline
  &   &   & & & \\
\multicolumn{6}{c}{\textbf{IGR J17315$-$3221}}\\
   &   &   &  & &  \\
\#1$^{a}$        &  1.24   &            --       &    $1.08\pm0.09$  &  
95.8/88  & $1.95\pm0.05$ \\
  &   &   &  & & \\
  \hline
  &   &  & & & \\
\multicolumn{6}{c}{\textbf{IGR J17449$-$3037}}\\
&   &   & & &  \\
\multicolumn{6}{c}{\textbf{obs1}}\\
  &   &  & &   &   \\
\#1        &  0.968   & $9.39^{+6.88}_{-7.02}$     &  $1.25^{+2.09}_{-1.48}$ &  
12.4/12  & $1.73\pm0.22$ \\
  &   &  & & & \\
  \multicolumn{6}{c}{\textbf{obs2}}\\
    &   & &    &   &    \\
\#1        &  0.968   &   $7.56^{+4.59}_{-3.26}$     &  $0.93^{+0.73}_{-0.62}$ &
21.0/21 & $1.35\pm0.08$ \\
    & &  &   &   &    \\
  \multicolumn{6}{c}{\textbf{obs3}}\\
     & &  &   &   &   \\
   \#1        &  0.968 & $3.01^{+2.30}_{-1.76}$     &  $0.19^{+0.51}_{-0.47}$   &
   19.5/23  &   $1.49\pm0.09$     \\
   &  &   &   &   &   \\
   \hline
   &  &  &  &  &  \\
     \multicolumn{6}{c}{\textbf{IGR J17596$-$2315}}\\
&   &   & & & \\
\#1        &  1.23   &           --          &  $0.15^{+0.43}_{-0.49}$ &
9.1/6 & $1.52\pm0.15$\\
    &   &     &   &   &    \\ 
\hline
\hline
\end{tabular}
\begin{list}{}{}
\item $^{a}$: For this source the best-fit model includes a narrow Gaussian line: energy 
centroid $E_{c}=6.36\pm0.07$ keV and $EW = 208^{+120}_{-116}$ eV.
\end{list}
\end{center}
\end{table*}

\subsection{IGR J17449$-$3037}

\begin{figure}
   \centering
   \includegraphics[width=0.5\textwidth, angle=0]{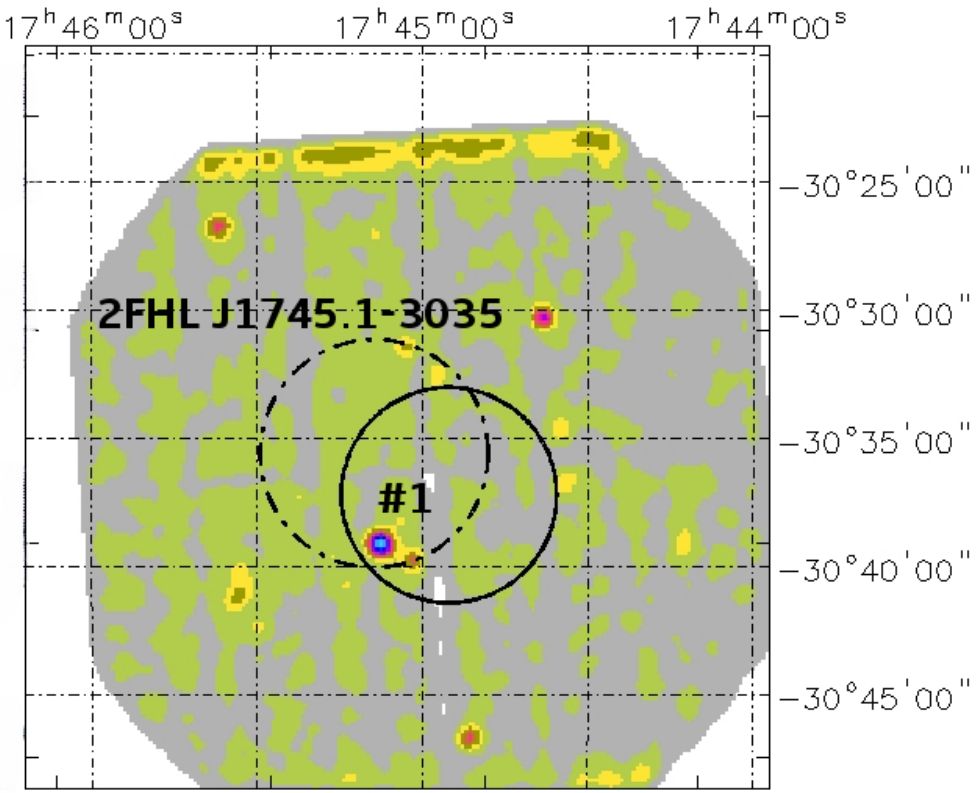}
   \caption{\emph{XMM-Newton} pn 0.2--12 keV image of the region surrounding IGR J17449$-$3037. 
There is only one source within both the black circle and the black-dashed-dotted error ellipse, 
which represent the positional uncertainty of IGR J17449$-$3037 and 2FHL J1745.1$-$3035, respectively.} 
              \label{fig7}%
    \end{figure}

Also IGR J17449$-$3037 is a new \emph{INTEGRAL} detection \citep{Krivonos2022}, close to the 
Galactic centre region. The source is particularly interesting since it lies in a region 
populated by GeV/TeV sources; in particular, the IBIS error circle crosses the positional 
uncertainty of the \emph{Fermi} source 2FHL J1745.1$-$3035 (see the \emph{XMM-Newton} view of the 
sky region around IGR J17449$-$3027 shown in Figure~\ref{fig7}) reported in the catalogue of 
\citet{Ackermann2016}, which lists objects detected above 50 GeV. Given the overlap of the two 
positional errors and the detection of a single X-ray source within both, we assume that the IBIS 
and LAT objects are associated and that the \emph{XMM-Newton} detection is their low-energy 
counterpart.

This hard \emph{Fermi} source is itself spatially coincident with the extended source HESS 
J1745$-$303, which may contain up to three different sources \citep{Aharonian2008}; indeed, the 
position of 2FHL J1745.1$-$3035 is compatible with TeV emission region C (the second brightest 
region in the complex, see Figure 1 in \citealt{Aharonian2008}). However, the nature of the 
source is probably more complex because 2FHL J1745.1$-$3035 is slightly brighter at 1 TeV than 
the entire H.E.S.S. region and has also a harder spectrum\footnote{2FHL J1745.14$-$3035 is one of 
the 3 hardest sources in the 2FHL catalogue, implying a high-energy SED peak in the TeV band.} 
(spectral index of $1.25\pm0.38$ estimated by \emph{Fermi} versus $2.17\pm0.11$ as measured by 
H.E.S.S.). Besides, the situation is complicated by emission at lower GeV energies and by the 
vicinity of the Galactic centre. For these reasons, we only consider valid the association 
between IGR J17449$-$3037 and 2FHL J1745.1$-$3035 for the remainder of the discussion. Although 
the source had been monitored by XRT on various occasions (see Table~\ref{tabA1}), it was never 
detected, thus implying a flux level below $2\times10^{-13}$ erg cm$^{-2}$ s$^{-1}$ over the 
0.2--12 keV waveband. When all these observations are summed together a marginal detection at 
2.4$\sigma$ is obtained: the source position is at R.A.(J2000) = $17^{\rm h}45^{\rm m}07^{\rm 
s}.20$, Dec.(J2000) = $-$$30^\circ39^{\prime}02^{\prime \prime}.30$ (error radius of 6 
arcseconds); the average source flux results to be $1.8\times10^{-13}$ erg cm$^{-2}$ s$^{-1}$ in 
the 0.2--12 keV band, confirming the lack of detection in single XRT snapshots performed over the 
period 2012--2021. Fortunately, this sky region was also monitored by \emph{XMM-Newton} and, 
luckily, the source was detected over at least 4 occasions: 3 times during individual pointings 
on March 21, 2001, on April 3, 2017, and on March 16, 2021 (see Table~\ref{tabA1}) and one time 
during a slew of the satellite performed on March 9, 2014. Archival data indicate a change in 
flux during \emph{XMM-Newton} monitoring that we investigated by analysing the source behaviour 
during each snapshot. As shown in Table~\ref{tab3}, the X-ray spectrum, with respect to the basic 
model, requires extra absorption and is characterised by a flat photon index. The source was 
detected at a 0.2--12 keV flux level of 
$(1.73\pm0.22)\times10^{-12}$, 
$(1.35\pm0.08)\times10^{-12}$, and $(1.49\pm0.09)\times10^{-12}$ erg cm$^{-2}$ s$^{-1}$
during the individual pointing 
(see Table~\ref{tab3}).

If we also take into account the 0.2--12 keV flux estimate of $(0.73\pm0.39)\times10^{-12}$ erg 
cm$^{-2}$ s$^{-1}$ obtained during \emph{XMM-Newton} slew, we assess a source variability by a 
factor of 2.4 between individual \emph{XMM-Newton} observations and by a factor of $\sim$10 
comparing to XRT pointings. We also note that these changes can occur on short time scale, as 
evident from a comparison between the 2021 \emph{XMM-Newton} and XRT measurements (by a factor of 
over 10 over less than a month time scale). \emph{XMM-Newton} also allows a better positioning of 
the source as reported in Table~\ref{tab1}, with an accuracy at 1-arcsecond level.

Due to the \emph{XMM-Newton} reduced positional uncertainty, we are also able to find a unique 
counterpart in the VIRAC (VVV Infrared Astrometric Catalogue, \citealt{Smith2018}) source 
195465789 located at R.A.(J2000) = $17^{\rm h}45^{\rm m}08^{\rm s}.0180$, Dec.(J2000) = 
$-$$30^\circ39^{\prime}04^{\prime \prime}.993$, having $Ks$ magnitude of $16.707\pm0.031$ and a 
total proper motion of $199.88\pm81.14$ mas/yr. Hence, we have a source that emits from few keV 
to GeV energies, it is variable in X-rays, quite hard in GeV, and must be of Galactic nature 
since the VIRAC source has a proper motion. We also note that hard sources in the 2FHL catalogue 
(with $\Gamma <2$) tend to be associated with Galactic objects, thus confirming that we are 
dealing with a close-by source and also with an efficient particle accelerator. Among Galactic 
sources that emit above 50 GeV, many are Supernovae remnants (SNRs), Pulsar Wind Nebulae (PWNs) 
and PWN/SNRs complexes with a small contribution (only 5 objects) of binary systems 
\citep{Ackermann2016}. SNRs tend to have softer spectra and are generally not variable in X-rays, 
while strong variability can be a distinctive properties of binary systems and less of PWNs. Also 
we note that no radio emission is detected in the direction of the source, which is unlikely for 
a SNR or even a PWN. Thus, we conclude that this may be a new GeV (maybe TeV) emitter possibly 
associated with a binary system or, at a lower level, a PWN.

Unfortunately, the location of the source close to the Galactic plane does not favour optical or 
IR observations (although we note that the VIRAC source is quite bright in $Ks$ magnitude), but 
\emph{Chandra} high-resolution data could allow one to confirm or exclude a PWN association by 
finding whether extended X-ray emission is observed or not.

\subsection{IGR J17596$-$2315}

This new source, reported by \citet{Krivonos2022}, is located in the Sagittarius region. It is 
another case where a \emph{Fermi} source, 4FGL J1759.5$-$2312, lies nearby. This GeV source is 
still unclassified and poorly studied: its emission can be described by a power law with a photon 
index of $2.4\pm0.1$ and a 0.1--100 GeV flux of $2.1\times10^{-11}$ erg cm$^{-2}$ s$^{-1}$ 
\citep{Abdollahi2022}. The only association reported is with a radio source belonging to the NRAO 
VLA Sky Survey (NVSS, \citealt{Condon1998}), namely NVSS J175948$-$230944, of unknown type.

\begin{figure}
   \centering
 \includegraphics[width=0.5\textwidth, angle=0]{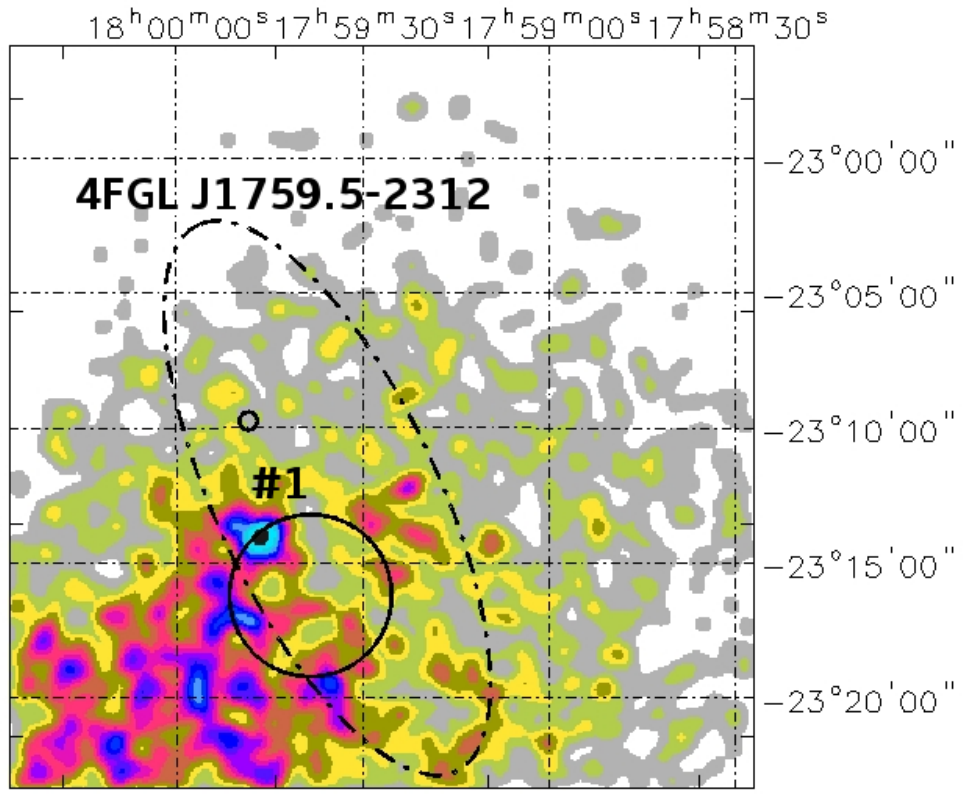}
   \caption{XRT 0.3--10 keV image of the region surrounding IGR J17596$-$2315. 
The only XRT detection lies at the border of the 90\% IBIS positional uncertainty (black circle) 
and is located within the error ellipse of 4FGL J1759.5$-$2312 (black-dashed-dotted ellipse). 
The smaller black circle depicts the position of the radio source NVSS J175948$-$230944, suggested 
to be the counterpart to the \emph{Fermi} object.}
              \label{fig8}%
    \end{figure}

The 0.3--10 keV XRT image of the region surrounding IGR J17596$-$2315, displayed in 
Figure~\ref{fig8}, shows the presence of an X-ray source located at the border of the 90\% IBIS 
error circle that also lies within the positional uncertainty of the \emph{Fermi} source 4FGL 
J1759.5$-$2312; interestingly, this X-ray source is not associated with the radio source NVSS 
J175948$-$230944, suggested as possible counterpart to the \emph{Fermi} emitter. Moreover, the 
XRT image indicates the presence of diffuse emission around the source of interest here, which is 
probably part of the SNR W28 located to the south and is large enough to extend up to the region 
pointed by XRT. Despite the presence of this emission, the overlapping of IBIS and LAT positional 
uncertainties, as well as the presence of a relatively bright XRT source inside both, points to a 
likely association between the 3 emitters.

Checking individual XRT pointings, we found that the XRT source was visible only on two occasions 
(at around 2.8$\sigma$), namely on February and August 2014 and undetected in other occasions 
(probably due to the low exposure times of the pointings). Unfortunately, because of the low 
statistical quality of the X-ray data, we could not characterise the source emission in each 
epoch. Therefore, we summed together the data and perform the spectral analysis of the average 
spectrum. By fitting the data with our basic model and fixing the photon index to 1.8, we found a 
2--10 keV flux of $\sim$$8.6\times10^{-13}$ erg cm$^{-2}$ s$^{-1}$ (see also Table~\ref{tab2}).

The source was also observed by \emph{XMM-Newton} on March 19--20, 2003; since we did not find 
evidence for flux variability in the data, we stack together the observations and performed the 
analysis of the average spectrum. Our basic model yields a good fit to the data (see 
Table~\ref{tab3}), providing a flat photon index ($\Gamma \sim 0.2$) and a 0.2--12 keV flux of 
$\sim$$1.5\times10^{-12}$ erg cm$^{-2}$ s$^{-1}$. We then applied the \emph{XMM-Newton} best-fit 
model to the XRT data, finding that in a similar energy band the flux is comparable 
($\sim$$1.2\times10^{-12}$ erg cm$^{-2}$ s$^{-1}$), thus suggesting no flux variation for over a 
decade. The \emph{XMM-Newton} data are also useful to restrict the positional uncertainty of the 
source, which reduces to 2 arcseconds for R.A.(J2000) = $17^{\rm h}59^{\rm m}46^{\rm s}.495$, 
Dec.(J2000) = $-$$23^\circ13^{\prime}58^{\prime \prime}.39$.

The \emph{XMM-Newton} data confirm the presence of diffuse emission suggested by XRT and raise 
the possibility that we are looking at a bright X-ray spot in the outskirts of the SNR W28.

Within the \emph{XMM-Newton} error circle we find the VIRAC \citep{Smith2018} source 223122211, 
which is characterised by a proper motion of $1.24\pm0.53$ mas/yr and IR magnitudes 
$J=15.919\pm0.016$, $H=14.363\pm0.014$ and $Ks=13.561\pm0.015$. The source is also visible in the 
Panoramic Survey Telescope \& Rapid Response System (Pan-STARRS) catalogue \citep{Chambers2016}, 
with an $I$ magnitude of $21.43\pm0.16$; finally, it is also reported in the VPHAS survey 
\citep{Drew2014} with a photometric H$\alpha$ magnitude of $20.76\pm0.13$. The observation of 
proper motion clearly points to a Galactic object, most likely a star of some type (with a 
probability $\ge$ 90\%, according to the UKIDSS-DR6 Galactic Plane Survey, \citealt{Lucas2008}). 
On the other hand, the near-IR data allow us to estimate the index $Q= (J-H)-1.7(H-K)$, which 
provides a way to separate early-type from late-type stars \citep{Reig2016}. While the latter are 
mostly concentrated around values of $Q=0.4-0.5$ (which correspond to spectral types K to M), the 
early-type objects typically display $Q < 0$ \citep{Reig2016}. In this case, $Q=0.17\pm0.03$ 
suggests an early-type object.

\begin{figure}
   \centering
    \includegraphics[width=0.45\textwidth, angle=0]{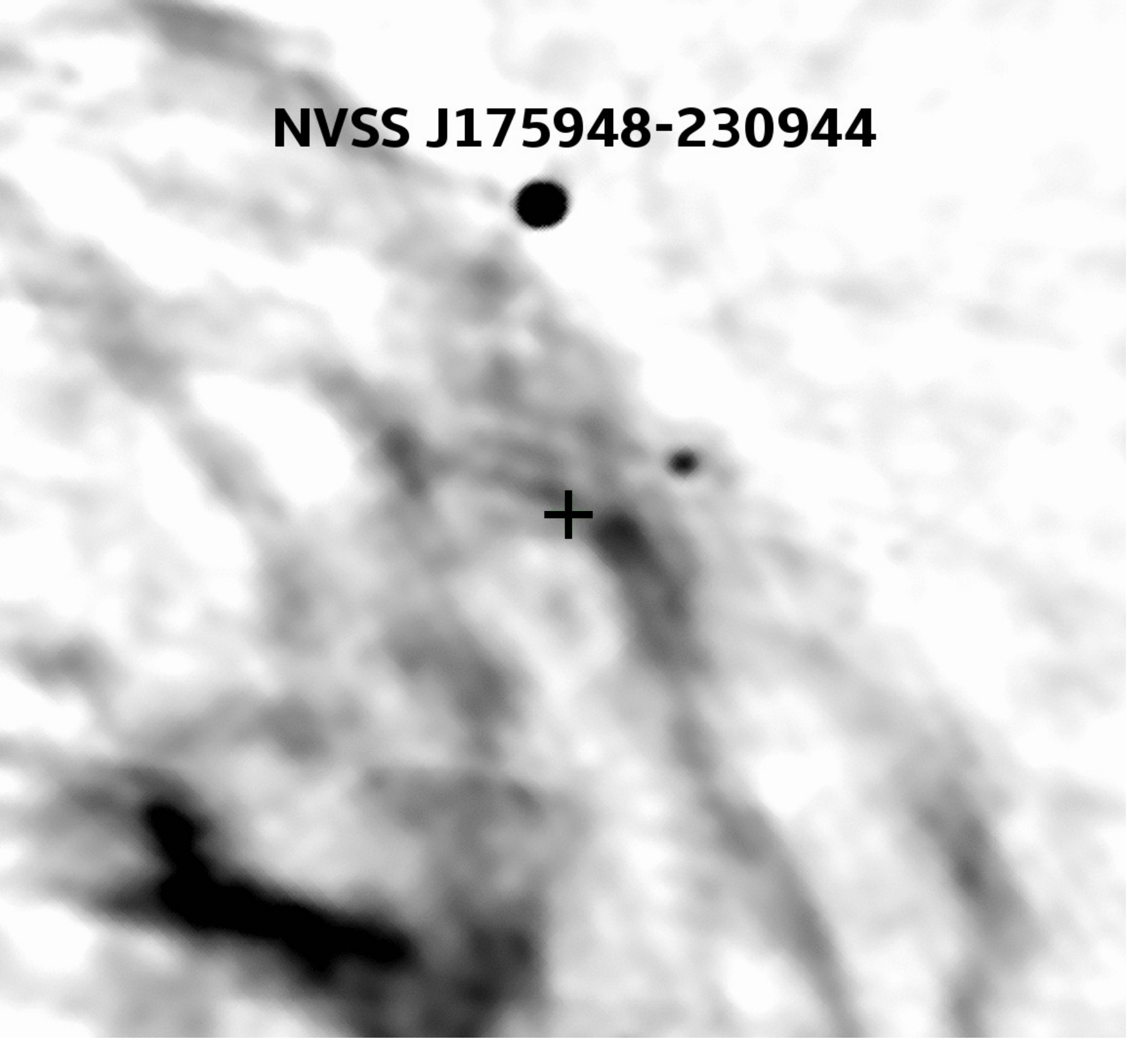}
   \caption{RACS radio image of the region surrounding IGR J17596$-$2315. 
The black cross depicts the position of the source detected in X-rays, which is located at 
   R.A.(J2000) = $17^{\rm h}59^{\rm m}46^{\rm s}.56$ and 
Dec.(J2000) = $-$$23^\circ13^{\prime}56^{\prime \prime}.43$. Also highlighted is the radio source 
   NVSS J175948$-$230944 that was proposed as the counterpart to the \emph{Fermi} source
   4FGL J1759.5$-$2312.}
              \label{fig9}%
    \end{figure}

\begin{figure}
   \centering
     \includegraphics[width=0.5\textwidth, angle=0]{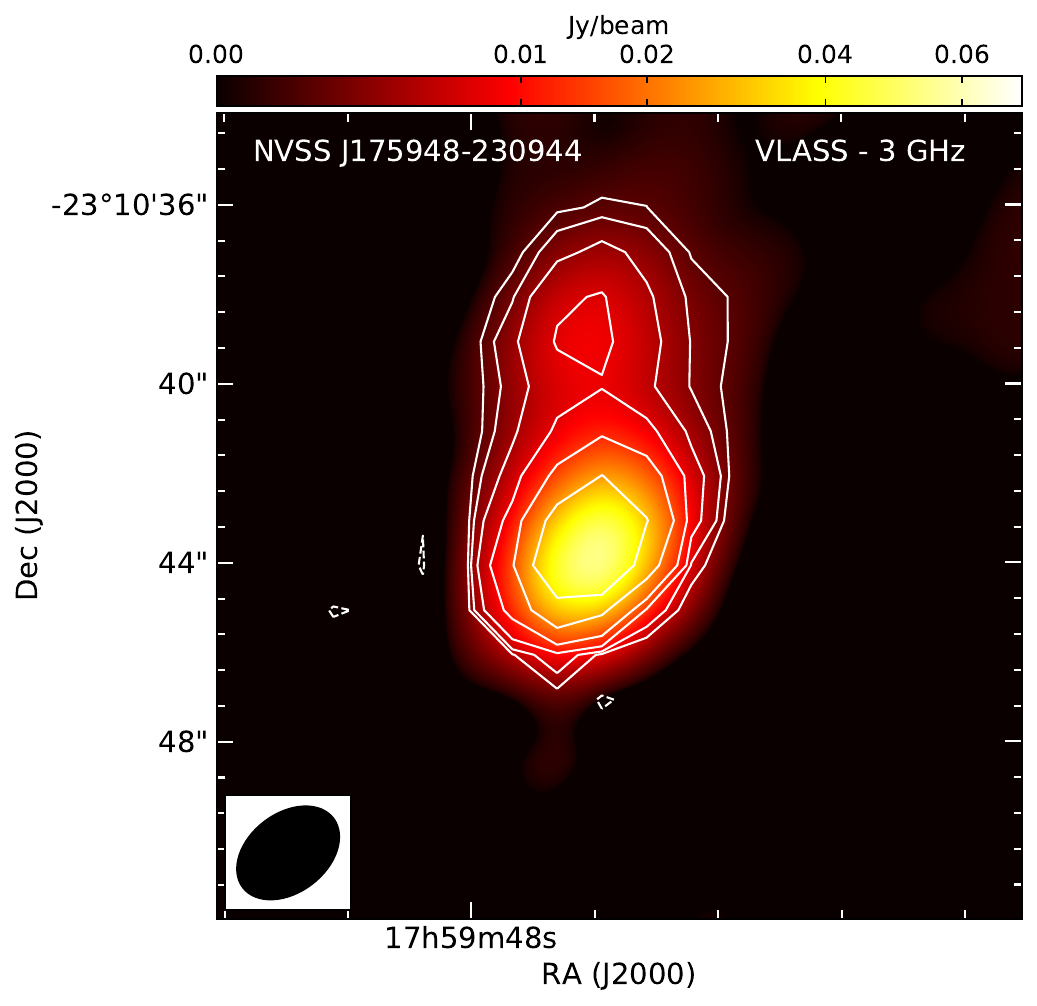}
   \caption{VLASS image of NVSS J175948$-$230944.}
              \label{fig10}
    \end{figure}

No radio emission is detected from this X-ray source (see Table~\ref{tabB1}),
which is however located at the outskirt of 
the SNR W28 and close to a radio filament as clearly visible in the Rapid ASKAP Continuum Survey 
(RACS, \citealt{Lacy2020}) image of the region (see Figure~\ref{fig9}); unfortunately, it is 
difficult with present data to assess whether the source is part of the remnant structure or 
simply an X-ray emitting object behind it.

The bright radio source located north of the IBIS source is NVSS J175948$-$230944, suggested to 
be the counterpart to the \emph{Fermi} source 4FGL J1759.5$-$2312. In reality, NVSS did not 
pinpoint the correct position of the source, better localised in the catalogues of compact radio 
sources in the Galactic plane by \citet{White2005}, where it is listed with a 1.4 and 5 GHz flux 
of $53.6\pm0.2$ and $157.5\pm2.5$ mJy, respectively. The source is also visible in images by RACS 
\citep{Lacy2020}, with a 0.88 GHz flux of $330\pm21$ mJy and by the Very Large Array Sky Survey 
(VLASS, \citealt{McConnell2020}), with a 3 GHz flux of $118\pm12$ mJy. The RACS position position 
is at R.A.(J2000) = $17^{\rm h}59^{\rm m}47^{\rm s}.77$, Dec.(J2000) = 
$-$$23^\circ10^{\prime}41^{\prime \prime}.75$ (2.5 arcseconds uncertainty); the morphology is 
that of a complex source showing a bright core with extended emission on one side as evident from 
the VLASS image shown in Figure~\ref{fig10}. The source is pretty weak in optical with a 
Pan-STARRS $I$ magnitude of $21.36\pm0.09$ and relatively bright in IR with UKIDSS-DR6 $J$ and 
$H$ magnitudes of $15.455\pm0.006$ and $13.975\pm0.04$, respectively. This source could be a 
blazar seen behind our Galaxy and thus be a possible counterpart to the \emph{Fermi} object. On 
the other hand, the presence of an X-ray source with emission up to 100 keV indicates that the 
association with 4FGL J1759.5$-$2312 is very likely.

\subsection{IGR J18006$-$3426}

Also IGR J18006$-$3426 is a new addition to the \emph{INTEGRAL} list of high-energy emitters by 
\citet{Krivonos2022}, who associated the source with the XRT catalogued object 2SXPS 
J180050.6$-$342322 (error radius of 4.9 arcseconds, see \citealt{Evans2020}) discovered thanks to 
a previous pointing performed on February, 2013; however, the reported 0.3--10 keV flux was quite 
high ($1.6\times10^{-11}$ erg cm$^{-2}$ s$^{-1}$) for a source seen at only few sigma c.l., as 
typically observed with XRT. Reanalysing these data we found the source to be barely detected 
(3.5$\sigma$ c.l. overall) at the border of the field of view (roughly 12 arcminutes from the 
pointing) and only in one of the two observations (that on February 8, 2013); its observed flux, 
estimating by fitting the data with our basic model, was also well below the reported one (around 
$4.6\times10^{-12}$ erg cm$^{-2}$ s$^{-1}$ in the 0.3--10 keV energy range), thus casting some 
doubts on its real presence and/or true X-ray properties.

\begin{figure}
\centering
  \includegraphics[width=0.5\textwidth, angle=0]{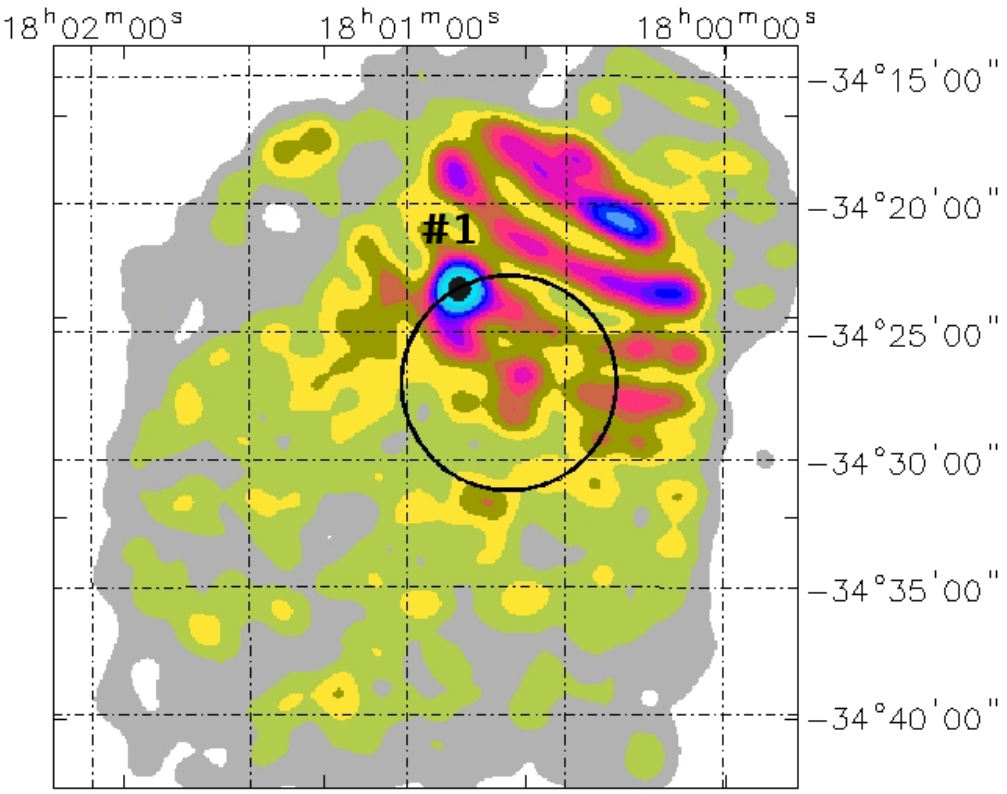}
   \caption{XRT 0.3--10 keV image of the region surrounding IGR J18006$-$3426. The only X-ray 
detection is located at the border of the 90\% IBIS positional uncertainty (black circle).} 
              \label{fig11}
    \end{figure}

For these reasons, we triggered a \emph{Swift}/XRT ToO observation at the position of the IBIS source, 
which 
was performed on February 6, 2023 (see Table~\ref{tabA1}). The source detection is confirmed at 
the location reported in Table~\ref{tab1}, with a 0.3--10 keV flux of $\sim$$3\times10^{-12}$ erg 
cm$^{-2}$ s$^{-1}$ evaluated adopting our basic model, which is slightly below our 2013 estimate. 
However, given the large uncertainties, we assume that the source did not vary over this long 
time span and therefore summed the two pointings together to enhance the signal-to-noise-ratio.  
The resulting image is shown in Figure~\ref{fig11}, where the unique X-ray counterpart is visible 
within the IBIS positional uncertainty of 4.2 arcminutes.  The average spectrum is well described 
by the basic model and it is characterised by a flat spectrum ($\Gamma \sim 1.1$) (see 
Table~\ref{tab2}).

Despite the XRT positional uncertainty being too large to pinpoint a unique optical/IR 
counterpart, we note that a radio source, VLASS1QLCIR J180050.84$-$342322.0 is detected by VLASS 
\citep{McConnell2020} within the soft X-ray error circle; its flux is 2.37$\pm$0.26 mJy at 3 GHz 
and its morphology is reported as compact (in reality, it is slightly resolved at the resolution 
of the instrument with a deconvolved size of $2.32\pm0.59\times0.52\pm0.44$ arcseconds). The 
source is also reported in the NVSS survey \citep{Condon1998}, with a 1.4 GHz flux of $2.9\pm0.6$ 
mJy, while in RACS \citep{Lacy2020} it is detected with a 0.88 GHz flux of $7.85\pm0.3$ mJy (see 
Table~\ref{tabB1}). While a comparison between VLASS and NVSS fluxes suggests a 
flat spectral index ($\alpha = -0.29$), that between RACS and VLASS/NVSS indicates instead an 
unphysical steep $\alpha$ value (from --0.96 up to --2.14); this probably hints at a change in 
flux at lower frequencies over a time span of around 30 years.

This source is also listed in the \emph{AllWISE} catalogue \citep{Cutri2014} with magnitudes 
$W1=13.386\pm0.085$, $W2=12.186\pm0.038$, $W3=8.960\pm0.035$ and $W4=6.358\pm0.072$; the source 
IR colours are therefore $W1-W2=1.2$ and $W2-W3=3.23$, which are typical of highly efficiently 
accreting AGN \citep{Secrest2015}. Using a third colour $W3-W4=2.6$ and the three-dimensional 
\emph{WISE} colour space discussed by \citet{Dabrusco2019} (see their Figure 6) it is possible to 
characterise further the source as a blazar and more specifically as a Flat Spectrum Radio Quasar 
(FSRQ); this is confirmed by the listing of the source in the \emph{Gaia} large amplitude 
variables \citep{Mowlavi2021} and VIVACE (VIrac VAriable Classification Ensemble, 
\citealt{Molnar2022}) catalogues as a long-period variable. The slightly extended structure seen 
in VLASS could also suggests a core plus jet morphology. We thus conclude that the radio source 
is a blazar behind the Galactic plane.

Taking into account that a chance coincidence cannot be excluded in this crowded region and that 
other objects are present within the XRT positional uncertainty, we consider the association 
between the IBIS emitter and the VLASS source quite probable but not definitive until a more 
accurate X-ray position is provided. If this association is confirmed, then IGR J18006$-$3426 can 
be classified as a new blazar shining behind our Galaxy.

\subsection{IGR J19071$+$0716}

\begin{figure}
   \centering
     \includegraphics[width=0.5\textwidth, angle=0]{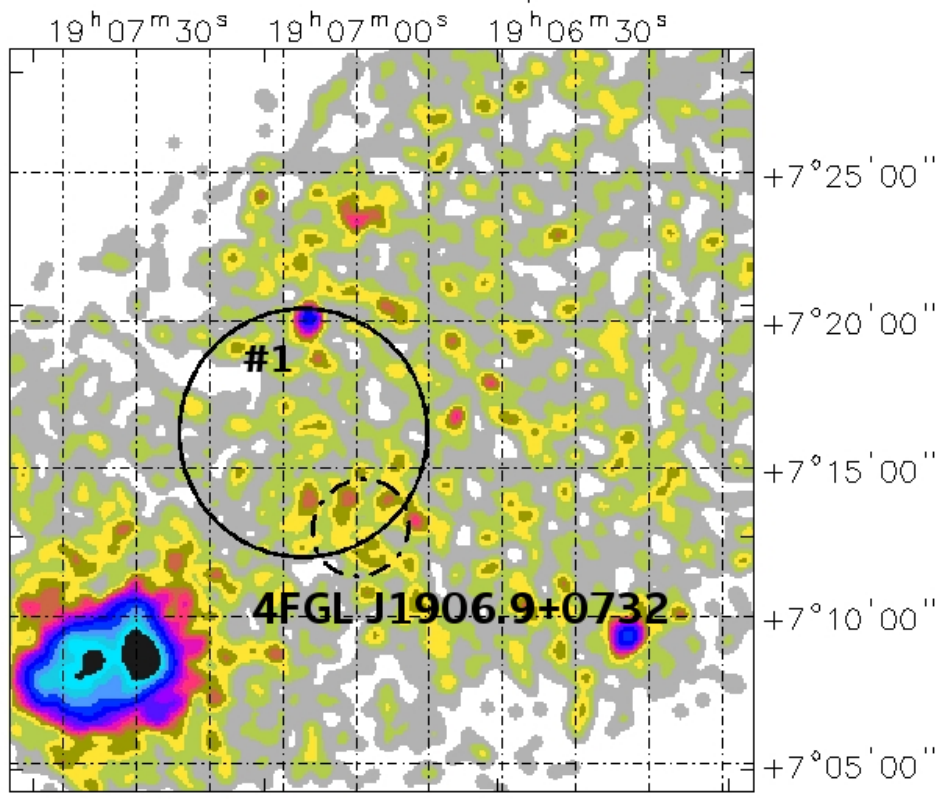}
   \caption{XRT 0.3--10 keV image of the region surrounding IGR J19071$+$0716. 
   The XRT detection lies at the border of the IBIS error circle (black circle). 
Also plotted is the positional uncertainty of the \emph{Fermi} source 4FGL J1906.9$+$0712
(black-dashed-dotted ellipse), which partially intersects the IBIS error circle.} 
              \label{fig12}%
    \end{figure}

This source, first reported in the \citet{Krivonos2017} list of \emph{INTEGRAL} sources and 
located in the Aquila region, is still unidentified in the latest catalogue by the same authors 
\citep{Krivonos2022}, where it is highlighted as a low signal-to-noise source ($<$4.5$\sigma$ 
c.l.). This may be due to the weak or variable nature of the source. Indeed, this sky region was 
observed several times by XRT (albeit with different exposures, see Table~\ref{tabA1}), but only 
in 2010 during a long pointing, and marginally in 2012, a single source was detected inside the 
IBIS error circle (see Figure~\ref{fig12}). These observations allow us to locate the source with 
higher precision and to restrict significantly the IBIS positional uncertainty (see 
Table~\ref{tab1}). Unfortunately, due to the low signal-to-noise ratio of the X-ray data, we 
could perform a reliable spectral analysis only for the longest pointing. The source spectrum can 
be approximated by a flat power law ($\Gamma \sim 1.5$) with a 2--10 keV flux of 
$\sim$$3.5\times10^{-13}$ erg cm$^{-2}$ s$^{-1}$ (see Table~\ref{tab2}). However, from the 
observation performed in 2012 we can only infer a 2--10 keV flux around $3.4\times10^{-13}$ erg 
cm$^{-2}$ s$^{-1}$, if we assume our basic model with photon index frozen to 1.8, thus indicating 
that the source did not change in flux from 2010 to 2012.

Despite the good positional uncertainty of the XRT data, the location of the source on the 
Galactic plane prevents us from pinpointing one single optical/IR counterpart. Fortunately, this 
sky region was also observed by \emph{Chandra} on February 18, 2018: a source (2CXO J190706.3$+$072001 
detected at 4.1$\sigma$ c.l.), compatible with the XRT one, is found at R.A.(J2000) = $19^{\rm 
h}07^{\rm m}06^{\rm s}.24$, Dec.(J2000) = $+$$07^\circ20^{\prime}01^{\prime \prime}.90$ (error 
radius of 1.8 arcseconds), while the 0.5--7 flux is $(5.5\pm1.7)\times10^{-14}$ erg cm$^{-2}$ 
s$^{-1}$.

The source was also detected during 4 \emph{XMM-Newton} measurements performed during different 
periods of time (on October 15, 2003, on April 23--25, 2004 and, most recently, on October 22, 
2018; \citealt{Webb2020}); over this period the 0.2--12 keV source flux changed from 
$(1-1.3)\times10^{-12}$ erg cm$^{-2}$ s$^{-1}$ (2003--2004) to $0.45\times10^{-12}$ erg cm$^{-2}$ 
s$^{-1}$ (2018); this last source state is similar to that observed by XRT on November 2010 and 
February 2012. Thus, the source flux decreased significantly (roughly by a factor $>$ 2) over a 
few years time scale.

Within the \emph{Chandra} positional uncertainty, we find only one IR object listed in the 
UKIDSS-DR6 Galactic Plane Survey \citep{Lucas2008} as UGPS J190706.32$+$072001.8, with magnitudes 
$J = 18.809\pm0.088$, $H=18.132\pm0.117$, and $K=17.415\pm0.129$. The index 
$Q=(J-H)-1.7(H-K)=-0.54\pm0.23$, estimated from the near-IR data, points to an early-type star if 
the source is Galactic \citep{Reig2016}.

Lacking other information, the nature of this counterpart remains elusive, although follow-up 
observations are now feasible thanks to the reduced X-ray positional uncertainty and the presence 
of an IR source within it.

We finally note that an unidentified \emph{Fermi} source, 4FGL J1906.9$+$0712/3FHL 
J1907.0$+$0713, lies close-by and intersects the IBIS error circle in its southern part (see 
Figure~\ref{fig12}). This \emph{Fermi} emitter is an interesting source because of a detected 
break in the $\gamma$-ray spectrum at around 130 MeV and a change in spectral slope to a value 
around 3 \citep{Abdollahi2022}, which is a strong indication that the \emph{Fermi} source is a 
cosmic ray accelerator; however, its properties are not typical of a SNR, a confirmed type of 
cosmic accelerator. Massive protostar outflows, stellar winds from runaway stars, colliding wind 
binaries, and young stellar clusters have also been considered as candidate sources 
\citep{Ergin2021}.

Unfortunately, the only X-ray source detected within the IBIS positional uncertainty is outside 
the \emph{Fermi} error ellipse, thus casting doubts on the possible association between the two 
high-energy emitters.

\subsection{IGR J19193$+$0754}

This \emph{INTEGRAL} source is most likely associated with 1RXS J191907.6$+$075921, a 
\emph{ROSAT} Bright source (0.1--2.4 keV flux of $1.5\times10^{-12}$ erg cm$^{-2}$ s$^{-1}$) 
located at a distance of 5.4 arcminutes and with a positional error of 13 arcseconds. As 
demonstrated by \citet{Stephen2006}, such an association is nearly secure and sometimes 
sufficient to restrict the X-ray source positional uncertainty, but in this case not enough to 
pinpoint a unique optical counterpart.

Despite this, 1RXS J191907.6$+$075921 was associated in the literature with TYC 
1042$-$2233$-$1/ASAS J191907$+$0759.3, a rotating variable star \citep{Haakonsen2009, Kiraga2012} 
with an average $V$ magnitude of 11.4 and a period of 3.16 days\footnote{see 
https://asas-sn.osu.edu/variables/abac3e8b-492d-51a7-8a8e-be996938828b for more information.}. 
According to the All Sky Automated Survey (ASAS, \citealt{Pojmanski2002}), the source is very 
close to us (around 300 parsec) and was classified as a Rotational Variable, i.e. a star 
that changes in its apparent brightness due to large spots on its surface. These stars are 
subdivided into several types like Ap stars, some Am stars, RS CVn stars, and BY Dra stars, most 
of which are generally not found to be strong emitters in the 20--100 keV.

\begin{figure}
   \centering
     \includegraphics[width=0.5\textwidth, angle=0]{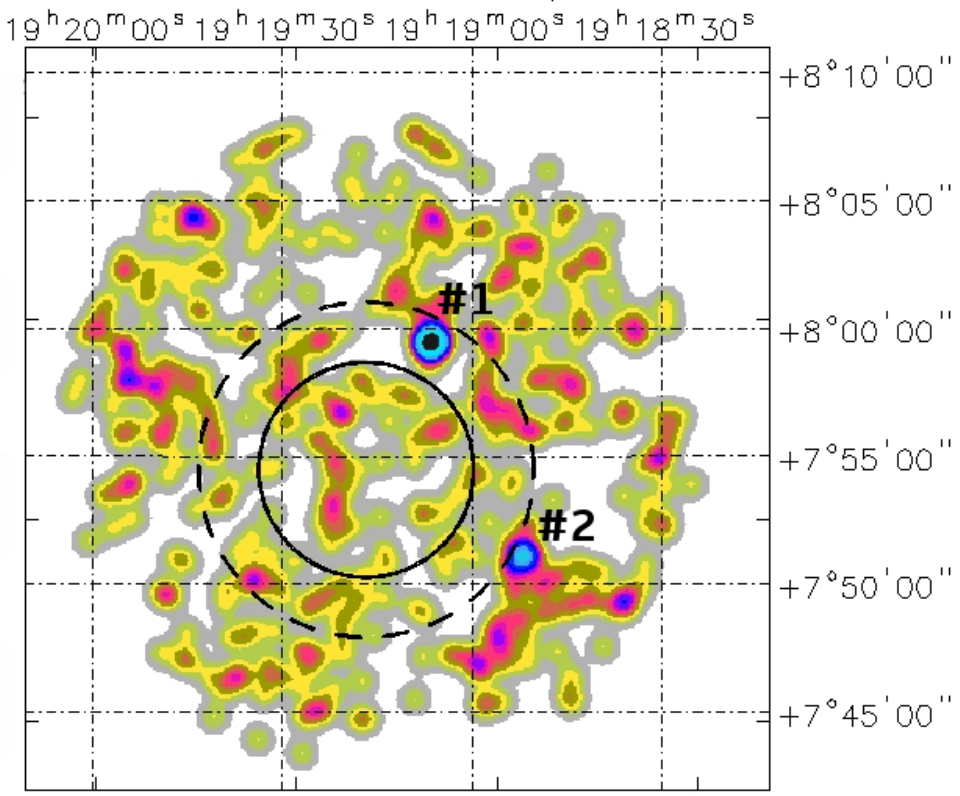}
   \caption{XRT 0.3--10 keV image of the region surrounding IGR J19193$+$0754. 
   The XRT detection, described in the text and labelled as source \#1, is outside the 90\% but 
inside the 99\% IBIS positional uncertainty (continuous and dashed black circles), respectively.}
              \label{fig13}%
    \end{figure}

Due to these uncertainties (relatively large soft X-ray error circle and unlikely X-ray emission 
from the proposed counterpart), we have requested a ToO observation of the source with 
\emph{Swift}/XRT, which was performed, on February 20, 2023 (see Table~\ref{tabA1}). The XRT 
image of the sky region is shown in Figure~\ref{fig13}, where two sources are detected inside 
(source \#1) and in proximity (source \#2) of the 99\% IBIS error circle. However, only source 
\#1 is of the right intensity to be a likely counterpart to the \emph{INTEGRAL} detection (see 
coordinates in Table~\ref{tab1}). The source location confirms that this is the \emph{ROSAT} 
Bright object but, unfortunately, the XRT exposure was not long enough to restrict considerably 
the positional uncertainty, which resulted to be only half the previous one in radius. The 
best-fit model is described only by a thermal component (MEKAL in {\sc XSPEC}), with a 
temperature around 0.6 keV. The 0.1--2.4 flux is $\sim$$1.4\times10^{-14}$ erg cm$^{-2}$ 
s$^{-1}$, thus indicating a variability by a factor of 100 if compared to the \emph{ROSAT} value.

The XRT error circle is still too large to pinpoint a single counterpart: for example \emph{Gaia} 
DR3 \citep{Brown2021} lists 6 objects, 5 of which are however extremely weak in optical, but one 
is quite bright. This is indeed the star associated in the literature with the \emph{ROSAT} 
detection, which is also reported as variable in various catalogues. The source is listed in the 
\emph{Gaia} DR3 catalogue (version 2022) with an apparent G magnitude of 10.87, a colour index 
$BP-RP=1.46$, and a distance of 325.9 pc; we also note that in \emph{Gaia} it was classified 
as a RS Canum Venaticorum type of variable star, again with a period of 3.16 days 
\citep[see][]{GaiaColl2022}\footnote{available at: 
https://vizier.cds.unistra.fr/viz-bin/VizieR?-source=I/354.}.

The \emph{Gaia} DR3 catalogue also quotes the source reddening corrected absolute $G$ magnitude 
and colour index ($M\_G=2.6$ and 1.1, respectively); these values are compatibles with those 
expected for RS Canum Venaticorum variable stars (see \citealt{Eyer2019}).

This type of stars is not typically found in the \emph{INTEGRAL} catalogue; a few of such objects 
were classified through optical spectroscopy \citep{Masetti2008,Masetti2012,Masetti2013}, while 
GT Mus \citep{Bird2016,Sguera2016} was discovered thanks to bright outbursts observed over the 
observing period (see also \citealt{Sasaki2021}). If this association is confirmed, this would be 
another such system caught by \emph{INTEGRAL}.

Therefore, if the association is correct, then the source must have been caught by XRT during a 
period of low flux, while presumably the \emph{INTEGRAL} detection could be related to one or 
more outburst events. Otherwise, this is not the correct association and a more refined 
positional error, such as one provided by \emph{Chandra}, is needed to pinpoint the right 
lower-energy counterpart.

\subsection{SWIFT J2221.6$+$5952}

\begin{figure}
   \centering
     \includegraphics[width=0.5\textwidth, angle=0]{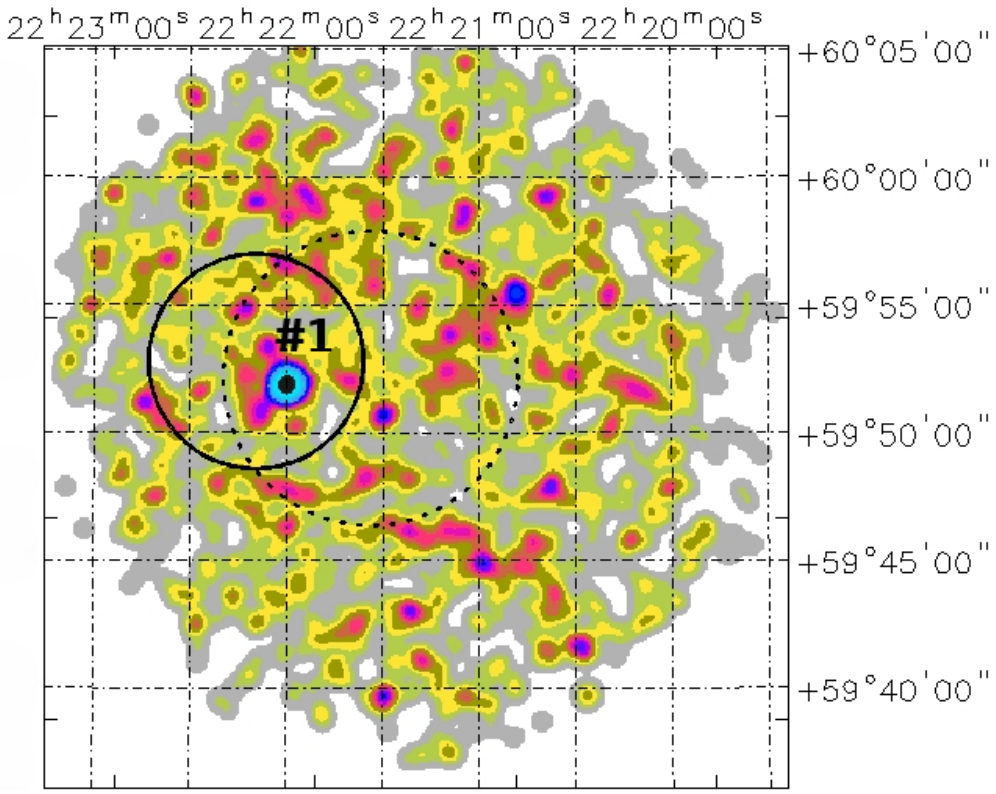}
   \caption{XRT 0.3--10 keV image of the region surrounding SWIFT J2221.6$+$5952. 
   The XRT detection lies inside both the IBIS and BAT error circles (continuous and dotted 
black circles, respectively).}
              \label{fig14}%
    \end{figure}

This source is another hard X-ray emitter that was listed for the first time in the 105-month 
\emph{Swift}/BAT catalogue \citep{Oh2018} and now reported also in the latest 
\emph{INTEGRAL}/IBIS survey \citep{Krivonos2022} as a source of unknown type (see 
Figure~\ref{fig14}). SWIFT J2221.6$+$5952 was observed by XRT on various occasions over the 
period July 2015 and October 2016 (see Table~\ref{tabA1}). From the analysis of each XRT 
measurements we can infer a 2--10 keV flux variability by a factor of $\sim$4 on a yearly time 
scale. However, due to the low statistical quality of the spectrum of each single pointing, we 
decided to stack all the observations together and perform the spectral analysis of the average 
spectrum, which can be modelled either with our basic model ($\Gamma \sim 0.9$) or with an 
absorbed power law ($N_{\rm{H(intr)}} \sim 0.8\times10^{22}$ cm$^{-2}$) if a canonical photon 
index of 1.8 is assumed (see Table~\ref{tab2}).

Within the soft X-ray positional uncertainty no counterpart is found in any waveband enquired, 
which makes this object a peculiar high-energy emitter, especially for a source that is located 
on the Galactic plane. In radio, the upper limit is below the mJy level at a few GHz 
(see Table~\ref{tabB1}), while in 
optical/IR the source is not visible to a level of 17 and 20 magnitudes in $J$ and in $B$ band, 
as inferred from the Two Micron All Sky Survey (2MASS, \citealt{Skrutskie2006}) and from the 
United States Naval Observatory (USNO–B1.0, \citealt{Monet2003}) catalogue, respectively.

The ratio of X-ray to optical fluxes is thus greater than 10, which may give some clues to the 
source nature. Indeed, although there is considerable dispersion, the most X-ray luminous 
(relative to optical) normal stars, normal galaxies, quasars, and BL Lac objects are known to 
have $log(f_{x}/f_{opt})$ of about --1, --0.25, 1.0, and 1.5, respectively (e.g., 
\citealt{Stocke1991}). This suggests an extragalactic nature for the source rather than a 
Galactic one.

SWIFT J2221.6$+$5952 is placed in the Cepheus constellation: this region is known to harbour 
intense star formation and various molecular clouds \citep{Kun2008}. It is also known for hiding 
background galaxies since \citet{Hubble1934} first noted that in this region the zone of 
avoidance extends towards high Galactic latitudes. It is therefore possible that a background 
radio quiet AGN remains hidden in this region at optical/IR wavelengths and is only visible at 
high energies.

Although difficult, given the source location, we nevertheless encourage further multi-band 
follow-up observations possibly with high-sensitivity telescopes in order to establish the nature 
of this source.

\section{Conclusions}

X-ray observations play a key role to reduce the positional uncertainty associated with 
unidentified, often newly discovered, soft $\gamma$-ray sources, though favouring the search of 
their likely counterpart/s, and finally unveiling their, often complex, nature.

On the other hand, because of the crowdedness of the Galactic plane at optical/IR wavelengths, we 
expect to find many objects falling within the X-ray positional uncertainty, thus making the 
identification of a single/double counterpart at this wavelengths difficult. Often a 
multi-waveband/multi-archive approach, as used in the present study, can help in the process, but 
unfortunately it is not always resolutive. Furthermore, variability also needs to be taken into 
account, since objects seen occasionally by high-energy instruments, may not be bright enough 
during follow-up at lower energies and vice versa. In view of this, it is useful to analyse data 
taken by different instruments at different epochs, such as done in this work where \emph{Swift}/XRT, 
\emph{XMM-Newton} and \emph{Chandra} observations were compared. Still despite this overall 
effort, the identification of high-energy sources on the Galactic plane remains a difficult task, 
as demonstrated in the present study.

The result of our work is summarised in Table~\ref{tab4}, where we provide, for each high-energy 
emitter, a tentative classification on the basis of the all information gathered. For IGR 
J16173$-$5023, IGR J17315$-$3221, IGR J17449$-$3037, IGR J17596$-$2315, and IGR J19071$+$0716, 
the more accurate position provided either by \emph{Swift}XRT, \emph{XMM-Newton}, or 
\emph{Chandra} enabled us to pinpoint a unique optical and/or IR counterpart. In the case of IGR 
J17327$-$4405, IGR J18006$-$3426, and IGR J19193$+$0754, the X-ray positional uncertainty is 
still too large to draw a firm conclusion about their optical/IR counterpart. For IGR 
J16267$-3303$, we found that the high-energy emission is most likely due to the contribution of 
two objects, which may be both classified as BL Lac. Our data also indicate that IGR 
J16173$-$5023 is likely a CV, IGR J17449$-$3037 a Galactic X-ray binary, IGR J18006$-$3426 a 
FSRQ-type Blazar, and IGR J19193$+$0754 a RS Canum Venaticorum system; IGR J17315$-$3221, IGR 
J17327$-$4405, IGR J17596$-$2315 and possibly IGR J19071$+$0716 are Galactic objects, although 
their specific optical class could not be assessed. In the case of SWIFT J2221.6$+$5952, no 
radio/optical/IR information is found for the X-ray counterpart, which makes this object peculiar 
also due to its location on the Galactic plane. Finally, two sources of the sample (IGR 
J17449$-$3037 and IGR J17596$-$2315) are positionally associated with the \emph{Fermi}/LAT 
unidentified sources 2FHL J1745.1$-$3035 and 4FGL J1759.5$-$2312, respectively; thus, the 
identification/classification of these GeV sources is a useful by-product of the present work.

\begin{table}
\small
\caption{Summary of the proposed counterparts.}
\label{tab4}
\begin{flushleft}
\begin{tabular}{lc}
\hline
\hline
Source & Type \\
\hline
\hline
IGR J16173$-$5023     &   Galactic (CV)   \\
IGR J16267$-$3303     &   Extragalactic (BL Lac)        \\
IGR J17315$-$3221     &   Galactic       \\
IGR J17327$-$4405     &   Galactic    \\
IGR J17449$-$3037     &   Galactic (X-ray binary?) \\
IGR J17596$-$2315     &   Galactic      \\
IGR J18006$-$3426     &   Blazar (FSRQ)  \\
IGR J19071$+$0716     &   Galactic ? \\
IGR J19193$+$0754     &   Galactic (RS Canum Venaticorum ?)        \\
SWIFT J2221.6$+$5952  &   Extragalactic ?          \\
\hline
\hline
\end{tabular}
\end{flushleft}
\end{table}

As a final remark, we wish to highlight that follow-up observations with current X-ray telescopes 
combined with multi-waveband studies (in particular optical/IR spectroscopy) are a powerful, but 
not always decisive, tool to assess the nature of unidentified high-energy emitter in the absence 
of better high-energy positioning capability.

\section*{Acknowledgements}

We acknowledge the anonymous referee for his/her valuable comments.
The authors acknowledge financial support from ASI under contract n. 2019-35-HH.0. This research 
has made use of data obtained from the VizieR catalog access tool and the SIMBAD database, which are 
both operated at CDS, Strasbourg, France; the 
NASA/IPAC Extragalactic Database (NED), which is operated by the Jet Propulsion Laboratory, 
California Institute of Technology, under contract with the National Aeronautics and Space 
Administration.
This work has made use of data from the European Space Agency (ESA) mission \emph{Gaia}
(https://www.cosmos.esa.int/gaia), processed by the \emph{Gaia}
Data Processing and Analysis Consortium (DPAC, https://
www.cosmos.esa.int/web/gaia/dpac/consortium). Funding
for the DPAC has been provided by national institutions, in
particular the institutions participating in the \emph{Gaia} Multilateral
Agreement.
This paper exploited data from \emph{XMM-Newton}, an ESA science mission with instruments 
and contributions directly funded by the ESA Member States and NASA.
We also acknowledge the use of public data from the \emph{Swift} data archive.

\clearpage

\appendix
\section{Log of the observations analysed in this paper}

In this Appendix we report the log of the observations used in this work.

 \setcounter{table}{0} 

\begin{table*}
\centering
 \caption{Log of the \emph{Swift}/XRT and \emph{XMM-Newton} observations used in this work. 
The exposure is expressed as net count rate.}
\footnotesize
\label{tabA1}
\begin{tabular}{llccc}
\hline
\hline
  IBIS source          &     ID        &      Obs date      &      Exposure    &  Instrument  \\
                       &               &                    &        (sec)     &              \\
 \hline
 \hline
                       &               &                    &                  &        \\
    IGR J16173$-$5023  &  00042853001  &  May 31, 2012      &      489         &  XRT   \\
                       &  00042853002  &  Jun 01, 2012      &      312         &  XRT  \\
                       &  00087349001  &  Apr 29, 2017      &     4810         &  XRT   \\
                       &  00041220001  &  Jan 14, 2019      &     1049         &  XRT   \\ 
                       &               &                    &                  &         \\
    \hline
                       &               &                    &                  &         \\
    IGR J16267$-$3303  &  00015479001  &  Jan 20, 2023      & 2068             &  XRT   \\
                       &  00096647001  &  Jan 22, 2023      & 1838             &  XRT   \\
                       &               &                    &                  &         \\
    \hline
                       &               &                    &                  &         \\  
    IGR J17315$-$3221  &  00043517001  &  Sep 01, 2012      &  247             &  XRT   \\
                       &  00034659002  &  Aug 08, 2016      &  1144            &  XRT   \\
                       &  00034659003  &  Aug 10, 2016      &  1603            &  XRT   \\
                       &  00034659004  &  Aug 11, 2016      &  969             &  XRT   \\
                       &  00034659005  &  Aug 12, 2016      &  1064            &  XRT   \\
                       &  00034659006  &  Aug 15, 2016      &  1066            &  XRT   \\
                       &  00034659007  &  Aug 16, 2016      &  966             &  XRT   \\
                       &  00034659009  &  Aug 18, 2016      &  1029            &  XRT   \\
                       &  00034659010  &  Aug 19, 2016      &  1083            &  XRT   \\
                       &  00034659011  &  Aug 20, 2016      &  1056            &  XRT   \\
                       &  0886040101   &  Sep 14, 2022      & 13730            &  XMM   \\
                       &               &                    &                  &         \\
    \hline
                       &               &                    &                  &        \\  
    IGR J17327$-$4405  &  00085657001  &  Apr 17, 2017      &  782             &  XRT  \\
                       &  00085657002  &  Oct 24, 2017      &  240             &  XRT  \\
                       &  00085657003  &  Mar 09, 2018      &  4595            &  XRT  \\
                       &  00087157001  &  Jan 30, 2019      &  967             &  XRT  \\  
                       &  00085657004  &  Jan 31, 2019      &  1086            &  XRT  \\ 
                       &  00087157002  &  Feb 01, 2019      &  974             &  XRT  \\
                       &  00085657005  &  Feb 27, 2019      &  1079            &  XRT  \\
                       &  00085657006  &  May 19, 2021      &  609             &  XRT  \\
                       &  00087157003  &  May 22, 2021      &  752             &  XRT  \\
                       &  00087157004  &  May 24, 2021      &  614             &  XRT  \\
                       &               &                    &                  &         \\
    \hline
                       &               &                    &                  &            \\
    IGR J17449$-$3037  &  00043600001  &  Apr 22, 2012      &  784             &  XRT        \\
                       &  00093484002  &  Apr 13, 2017      &  62              &  XRT    \\
                       &  00093484003  &  Apr 20, 2017      &  35              &  XRT     \\
                       &  00093484004  &  May 04, 2017      &  57              &  XRT     \\
                       &  00093484005  &  May 18, 2017      &  55              &  XRT       \\
                       &  00093484006  &  Jun 01, 2017      &  55              &  XRT       \\
                       &  00093484007  &  Jun 15, 2017      &  48              &  XRT       \\
                       &  00093488007  &  Jun 15, 2017      &  50              &  XRT        \\
                       &  00093484008  &  Jun 23, 2017      &  57              &  XRT       \\
                       &  00089069001  &  Oct 30, 2020      &  1628            &  XRT      \\
\hline
\end{tabular}
\end{table*}
\clearpage
\begin{table*}
\centering
\footnotesize
\begin{tabular}{llccc}
\hline
\hline
  IBIS source          &     ID        &	Obs date    &      Exposure    &  Instrument  \\
                       &               &                    &        (sec)     &              \\
\hline
\hline
                       &  00089069002  &  Apr 05, 2021      &  1600            &  XRT        \\
                       &  00089069003  &  Apr 07, 2021      &  1970            &  XRT        \\
                       &  0103261301   &  Mar 21, 2001      &  3864            &  XMM         \\
                       &  0782170601   &  Apr 03, 2017      &  5971            &  XMM        \\
                       &  0886010401   &  Mar 16, 2021      & 15360            &  XMM         \\
                       &               &                    &                  &                 \\
    \hline
                       &               &                    &                  &                \\
    IGR J17596$-$2315  &  00043799001  &  Feb 07, 2011      &  420             &  XRT    \\
                       &  00049688001  &  Feb 06, 2014      &  959             &  XRT   \\
                       &  00049688002  &  Aug 13, 2014      &  799             &  XRT   \\
                       &  00049688004  &  Nov 03, 2014      &  460             &  XRT         \\
                       &  00049688005  &  Feb 06, 2019      &  432             &  XRT        \\
                       &  0135742201   &  Mar 19, 2003      &  6866            &  XMM       \\
                       &  0135742301   &  Mar 20, 2003      &  6869            &  XMM     \\
                       &               &                    &                  &           \\
   \hline
                       &               &                    &                  &            \\
    IGR J18006$-$3426  &  00091419001  &  Feb 02, 2013      &  934             &  XRT         \\
                       &  00091419002  &  Feb 08, 2013      &  2225            &  XRT   \\
                       &  00015480001  &  Feb 06, 2023      &  1895            &  XRT        \\
                       &               &                    &                  &           \\
    \hline
                       &               &                    &                  &              \\
    IGR J19071$+$0716  &  00041804001  &  Nov 19, 2010      &  10673           &  XRT    \\
                       &  00044776001  &  Feb 17, 2012      &  1111            &  XRT   \\
                       &  00044775001  &  Mar 07, 2013      &  450             &  XRT         \\ 
                       &  00088696001  &  May 25, 2019      &  2095            &  XRT\\ 
                       &  03109633001  &  Nov 28, 2019      &  167             &  XRT     \\
                       &  03109633002  &  Nov 30, 2019      &  170             &  XRT       \\
                       &  03109633003  &  Dec 02, 2019      &  125             &  XRT       \\
                       &  03109633004  &  Dec 03, 2019      &  355             &  XRT      \\
                       &  03109633005  &  Feb 21, 2020      &  899             &  XRT \\
                       &  03109633006  &  Mar 02, 2020      &  812             &  XRT        \\
                       &  03109633007  &  Mar 09, 2020      &  981             &  XRT      \\ 
                       &               &                    &                  &         \\   
    \hline
                       &               &                    &                  &         \\
    IGR J19193$+$0754  &  00015470001  &  Feb 20, 2023      &  1910            &  XRT      \\
                       &               &                    &                  &     \\
    \hline
                       &               &                    &                  &          \\
    SWIFT J2221.6$+$5952 &  00085697001 & Jul 02, 2015      &  236             &  XRT     \\
                         &  00085697002 & J ul 27, 2015     & 1073             &  XRT    \\
                         &  00085697003 & Oct 08, 2015      &  607             &  XRT    \\
                         &  00085697004 & Oct 09, 2015      &  549             &  XRT    \\
                         &  00085697005 & Oct 25, 2015      &  356             &  XRT   \\
                         &  00085697006 & Apr 04, 2016      & 2447             &  XRT    \\
                         &  00085697007 & Apr 12, 2016      &  243             &  XRT  \\
                         &  00085697008 & Jul 08, 2016      &  514             &  XRT   \\
                         &  00085697009 & Jul 17, 2016      &  960             &  XRT   \\
                         &  00085697010 & Jul 25, 2016      &  228             &  XRT   \\
                         &  00085697011 & Aug 28, 2016      &  589             &  XRT   \\
                         &  00085697013 & Oct 23, 2016      &  900             &  XRT    \\
                         &              &                   &                  &         \\
  \hline
  \hline
\end{tabular}
\end{table*}
\clearpage

\section{Radio Data}

 \setcounter{table}{0}

\begin{table}[h!]
\small
\begin{flushleft}
\caption{Radio data of the likely counterparts to the IBIS sources studied in this work.}
\label{tabB1}
\begin{tabular}{lccc}
\hline
\hline
 Source &      RACS$^{a}$        &     NVSS                &    VLASS$^{a}$            \\
        &      (mJy)             &     (mJy)               &    (mJy)              \\
\hline
\hline
  \multicolumn{4}{c}{\textbf{IGR J16173$-$5023}}\\
\#1       &        $<6.0$               &       --               &        --            \\
\hline
  \multicolumn{4}{c}{\textbf{IGR J16267$-$3303}}\\
\#1 and \#2  &      $<0.79$             &       --               &      $<0.56$           \\
\hline
\multicolumn{4}{c}{\textbf{IGR J17315$-$3221}}\\
\#1        &            --             &        --             &         $<0.52$           \\
\hline
\multicolumn{4}{c}{\textbf{IGR J17327$-$4405}}\\
\#1        &            $<0.84$        &        --             &          --             \\
\hline
\multicolumn{4}{c}{\textbf{IGR J17449$-$3037}}\\
\#1        &           $<3.9$         &        --             &        $<0.45$             \\
 \hline
 \multicolumn{4}{c}{\textbf{IGR J17596$-$2315}}\\
\#1        &           --              &        --             &        $<0.61$             \\
\hline
  \multicolumn{4}{c}{\textbf{IGR J18006$-$3426}}\\
\#1        &       $7.85\pm0.30$       &     $2.90\pm0.60$     &       $2.39\pm0.26$      \\
\hline
\multicolumn{4}{c}{\textbf{IGR J19071$+$0716}}\\
\#1        &           $<3.0$            &        --             &         $<0.38$              \\
  \hline
  \multicolumn{4}{c}{\textbf{IGR J19193$+$0754}}\\
   \#1      &          $<1.5$            &        --             &        $<0.72$              \\
\hline
  \multicolumn{4}{c}{\textbf{SWIFT J2221.6$+$5952}}\\
\#1        &            --              &        --              &         $<0.35$            \\
\hline
\hline
\end{tabular}
\end{flushleft}
\begin{list}{}{}
\item $^{a}$: Upper limits are at 3$\sigma$ confidence level.
\end{list}
\end{table}

In order to search for radio emission for the entire sample, we consulted recent all-sky surveys 
like the VLA Sky Survey at 3 GHz (VLASS, \citealt{McConnell2020}), and the Rapid ASKAP Continuum 
Survey at 0.88 GHz (RACS, \citealt{Lacy2020}). The images for the fields of interest were 
downloaded and analysed with the {\tt CASA}\footnote{available at: https://casa.nrao.edu/} 
software \citep{CasaTeam2022}. Flux densities were extracted through Gaussian fitting, except for 
the only resolved source NVSS\,J175948-230944, for which we estimated the flux density in VLASS 
considering a region that included the entire source. In case of non-detection, the image RMS was 
estimated in a region centered on source coordinates. The radio flux densities obtained from all 
these surveys are listed in Table~\ref{tabB1}, which also specifies the database used. Inspection 
of Table~\ref{tabB1} indicates that most objects are not radio emitters with 3$\sigma$ flux upper 
limits of the order of the mJy.

As expected, the only source detected (IGR J18006$-$3426) is an AGN (see dedicated section for 
further details). On the other hand, neither of the possible AGN counterparts to IGR 
J16267$-$3303 is detected at a flux limit of a fraction of a mJy, thus making the \emph{WISE} 
colours the only signature of an extragalactic nature for this source.

\clearpage

\bibliographystyle{elsarticle-harv} 
\bibliography{INTEGRAL_unidentified_rev}






\end{document}